\documentclass[a4paper,fleqn,usenatbib]{mnras}

\usepackage{newtxtext,newtxmath}
\usepackage[T1]{fontenc}
\usepackage{graphicx}	
\usepackage{amsmath}	
\usepackage{amssymb}	
\usepackage{enumerate}
\usepackage{array}
\usepackage{xspace}
\usepackage{epstopdf}
\usepackage{soul}

\newcommand{\qdep}{$Q_{\rm dep}$\xspace}

\newcommand{\nickel}{$^{56}\rm Ni$\xspace}

\newcommand{\dmft}{$\Delta m_{15}$\xspace}

\title[SN Ia WLR II. Color]{Type Ia supernovae have two physical width-luminosity relations
	and they favor sub-Chandrasekhar and direct collision models - II. Color evolution}

\author[Wygoda et. al.]{
	Nahliel Wygoda,$^{1,2}$\thanks{E-mail: nahliel.wygoda@weizmann.ac.il}
	Yonatan Elbaz$^{2}$
	Boaz Katz$^{1}$
	\\
	$^{1}$Dept. of Particle Phys. \& Astrophys., Weizmann Institute of Science, Rehovot 76100, Israel\\
	$^{2}$Dept. of Physics, NRCN, Beer-Sheva 84190, Israel\\
}

\pubyear{2018}

\date{Accepted XXX. Received YYY; in original form ZZZ}

\begin{document}
\label{firstpage}
\pagerange{\pageref{firstpage}--\pageref{lastpage}}
\maketitle

\begin{abstract}
While the width-luminosity relation (WLR) among type Ia supernovae (slower is brighter) is one of the best studied properties of this type of events, its physical basis has not been identified convincingly. The 'luminosity' is known to be related to a clear physical quantity - the amount of $^{56}$Ni synthesized, but the 'width' has not been quantitatively linked yet to a physical time scale. 
We show that the recombination time of $^{56}$Fe and $^{56}$Co from doubly to singly ionized states causes the typical observed break in the color curve B-V due to a cliff in the mean opacities, and is a robust width measure of the light curve, which is insensitive to radiation transfer uncertainties. A simple photospheric model is shown to predict the recombination time to an accuracy of $\sim5$ days, allowing a quantitative understanding of the color WLR. Two physical times scales of the width luminosity relation are shown to be set by two column densities- the total column density which sets the gamma-ray escape time $t_0$ (previous Paper I) and the $^{56}$Ni column density which sets the recombination time (this Paper II). Central detonations of sub-$\rm M_{ch}$ WDs and direct WD collision models have gamma-ray escape times and recombination times which are consistent with observations across the luminosity range of type Ia's. Delayed detonation Chandrasekhar mass models have recombination times that are broadly consistent with observations, with tension at the bright end of the luminosity range and inconsistent gamma-ray escape times at the faint end.
\end{abstract}


\section{Introduction}
The width luminosity relation (WLR) of the light curves of type Ia supernovae (SNIa), stating that brighter SNIa have slower light curves, has long been established based on numerous observations \citep[e.g.][]{Pskovskii77,phil93,phil05}. It is important for correcting the absolute brightness of SNIa as standard candles for cosmological purposes, and it provides an important clue for identifying the unknown progenitors of SNIa. 
The luminosity of type Ia's is set by the amount of $^{56}$Ni \citep[e.g.][]{h+96,wkbs07} , but what physical properties of the ejecta sets the time scale (width)? 

It has been demonstrated that the recombination of iron group elements from doubly to single ionized states has a strong effect on the color evolution \citep{pi01,kw07,h+17} but not on the bolometric light-curves \citep{kw07} implying that more than one aspect of the ejecta is involved in setting the width of the light curves in different bands. One known key property of the ejecta is the total column density ($^{56}$Ni weighted and direction averaged) which sets the gamma-ray escape time $t_0$ \citep[e.g.][]{j99} that directly affects the bolometric light-curve evolution. In \cite{wygoda17a} (hereafter Paper I), it was shown that $t_0$ can be accurately and robustly inferred from the bolometric light curves of SNIa. In this work we show that the (pre-decayed) $^{56}$Ni column density is a second key property that sets the recombination time of the iron-group elements which directly affects the color evolution and can be extracted from observations.  

The color curves of type Ia have a striking break around 10-30 days after maximum light (see e.g. figure \ref{fig:burns_fig3}). It was recently realized by \cite{burns14} that the time of this break is an excellent WLR width parameter which is correlated with other known width parameters 
(such as the magnitude decline of the B-band within 15 days after max, \dmft, \citealt{phil93}) but is a better 'ordering parameter' for other characteristics of the light curves and in particular the properties of the infra-red bands. We use numerical radiation transfer simulations to show that this break is coincident with the recombination time of of the iron group elements. We demonstrate that its calculation is insensitive to radiation transfer uncertainties by comparing our calculations and previous calculations in the literature that employ different approximations. Moreover, we show that it can be analytically predicted by a simple photospheric emission model and is set by the (pre-decayed) $^{56}$Ni column density. 

We validate this simple model through a large sample of computed ejecta, and apply it to existing explosion models. In particular, collisions, sub-Chandrasekhar and delayed detonation Chandrasekhar mass models seem to have B-V breaks that are consistent with observations. The fact that delayed Chandrasekhar mass models have color breaks times that are consistent with observations but gamma-ray escape times that are not \citep[][see also \cite{kkd13}, \citealt{s+14}]{wygoda17a}, demonstrates that agreement of a model with only one aspect of the WLR provides little evidence for its validity. 

The structure of this paper is as follows: In section \S~\ref{sec:tbv_robust} we show that the calculation of the break time in the B-V curve  is insensitive to radiation transfer uncertainties by comparing the results of different calculations from the literature (and ours) involving different approximations. In section \S~\ref{sec:tbv_is_recombination} we show that the break time in the B-V color curve is coincident to the recombination time of the iron-group elements. In section \S~\ref{sec:tbv_predicted} we derive a simple analytic model to calculate the recombination time and show that it is set by the $^{56}$Ni column density. In section \S~\ref{sec:summary} we summarize the results and apply them to explosion models and observations of type Ia's.

\section{break time in b-v is robust}\label{sec:tbv_robust}
\begin{figure}
	\includegraphics[width=\columnwidth]{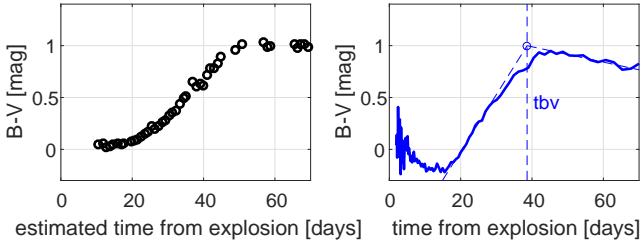}
	\caption{\label{fig:burns_fig3}
		Typical B-V light curve shape and break time. Left panel - SN2005M, from \protect\cite{burns14}, right panel -  simulation of a simple synthetic $^{56}\rm Ni$-C ejecta with $M(^{56}\rm Ni)=0.7M_{\odot}$, $M(^{12}\rm C)=0.45M_{\odot}$, high mixing, $t_0=35$ days (line 4 in table~\ref{table:ejecta_table}). 
	}
\end{figure}

The B-V color evolution of the typical type Ia SN2005M \citep[from][]{burns14} is shown in the left panel of figure \ref{fig:burns_fig3}. As can be seen there is a striking break in the curve (in this case around 35 days after maximum light, or approximately 53 days after explosion) where the evolution transitions from becoming redder with time to becoming bluer with time. It was shown in \cite{burns14} that this time scale can be used as an excellent width parameter for type Ia lightcurves. The right panel shows the result of a radiation transfer calculation of a synthetic ejecta with $0.7$ solar masses of $^{56}$Ni showing similar qualitative evolution. 

Similar to \cite{burns14}, we associate a precise time 'tbv' with the break as the point of intersection of two straight lines: one which is fitted to the rising portion of the B-V curve, and one to the constant or slightly declining portion which follows it as indicated in the right panel of figure \ref{fig:burns_fig3}. Note that this is slightly different than the precise choice adopted by \cite{burns14} which involved a fit to a continuous interpolation function. Note also that throughout this work, we measure the time of the break from \textbf{time of explosion} rather than from time of maximum B band luminosity. The physical meaning of the time of explosion is much more clear and while it is harder to infer observationally compared to the time of maximum B band luminosity, it can be measured to reasonable accuracy with sufficient early data.

\begin{figure*}
	\centering
	\includegraphics[width=\columnwidth]{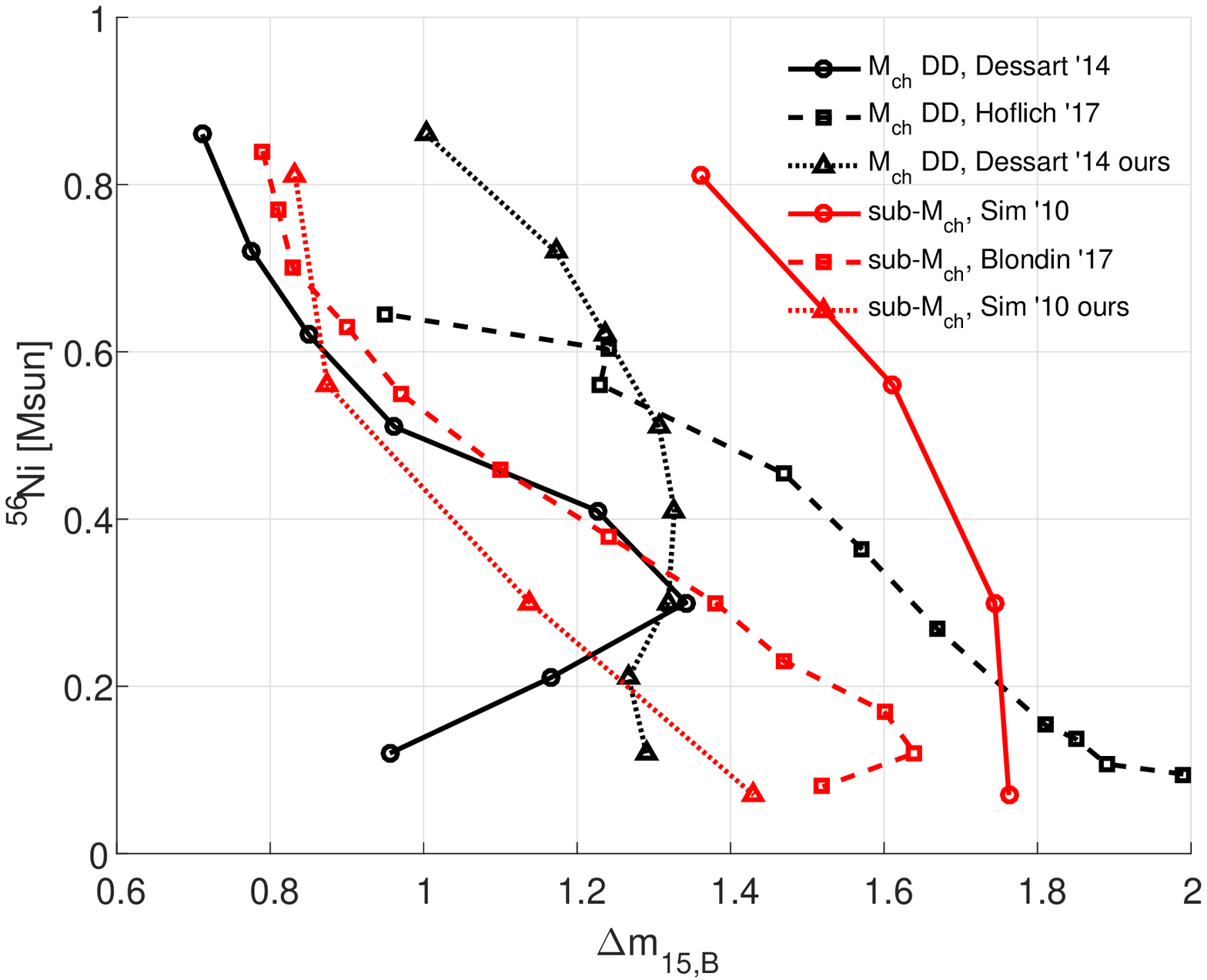}
	\hspace{2mm}
	\includegraphics[width=\columnwidth]{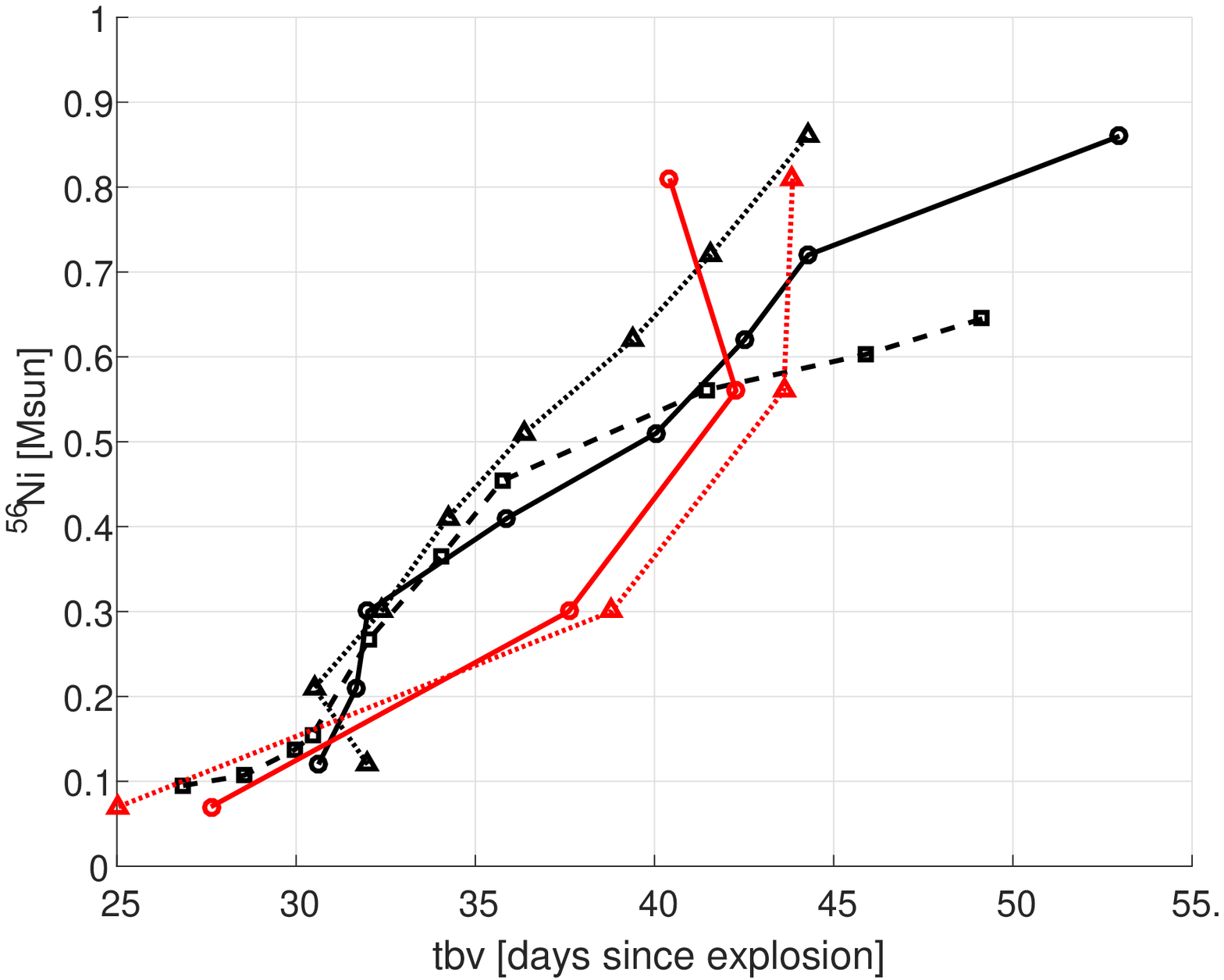}
	\caption{
		Simulated width luminosity relation (WLR)  with a standard width parameter, showing \nickel mass vs. \dmft (left panel), and a physically preferred WLR showing \nickel mass vs. the B-V break time tbv (right panel). 
		The relations are shown for two types of explosion models: Delayed detonations of $M_{\rm ch}$ WDs as computed by \protect\cite{dbhk14} (black solid), in our radiative transfer simulation of the same ejecta (black dotted), and by \protect\cite{h+17} (black dashed), and for central detonations of sub-$M_{\rm ch}$ WDs as computed by \protect\cite{s+10} (red solid) in our radiative transfer simulation of the same ejecta (red dotted), and by \protect\cite{b+17} (red dashed).} 
	\label{fig:WLRs}
\end{figure*}

Figure~\ref{fig:WLRs} shows two types of WLR obtained from simulations, one using the B-band post maximum decline rate \dmft (left panel) and one using tbv (right panel) for two types of physical models: Delayed detonations (DD) of Chandrasekhar mass WDs, and central detonation (CD) of sub-Chandrasekhar mass WDs. For the DD models, we show results from 3 different simulations: one by \cite{dbhk14}, a second one were we simulated the radiative transfer through the same ejecta in a different radiative transfer simulation (see ~\ref{sec:radiative_transfer}), and a third simulation by \cite{h+17}. For the CD model, we show the results of simulations done by \cite{s+10} as well as the results of our radiative transfer simulation for the same ejecta, and a third simulation by \cite{b+17} for a similar scenario.

As can be seen in the left panel, different calculations result in very large differences in \dmft for very similar ejecta (all sub-Chandra calculations have similar ejecta for given $^{56}$Ni mass, and all  Chandra calculations have similar ejecta for given $^{56}$Ni mass). The right panel, on the other hand, shows that differences in tbv between the different calculations are much less significant. 

Note that the different numerical codes use different approximations. In particular, the ionization and energy level population are treated differently. The calculations in \cite{dbhk14,b+17,h+17} solve detailed NLTE equations for the ionization and energy level populations. \cite{s+10} assume the level populations are in LTE but solve the ionization and recombination equations. Our calculation assumes level populations as well as the ionization structure is in LTE, but does not require the radiation field to be in equilibrium (see description in \S \ref{sec:radiative_transfer}) We note that fully LTE calculations seem to deviate significantly in the light curves shapes, including the break time, compared to the various previously mentioned approximations (see appendix)~\ref{sec:appendixA}.
In \cite{dbhk14,b+17,h+17} and our calculations, the electron temperature is solved for by requiring local thermal balance while in \citep{s+10} it is set by the local radiation temperature. While the different approximations lead to light curves that deviate substantially as evident in the left panel, the values of tbv are very similar for similar ejecta suggesting that it is insensitive to the uncertainties in the radiation transfer approximations.   

\section{break time in b-v occurs at recombination}\label{sec:tbv_is_recombination}

In this section, we study the origin of the typical break in the B-V curve using radiative transfer calculations of a wide range of ejecta.

\subsection{Radiative transfer simulations}\label{sec:radiative_transfer}
Radiation transfer is calculated using URILIGHT, a Monte-Carlo code based on the following approximations (described in \S\ref{sec:appendixA}, similar to the SEDONA program, \citealt{ktn06}): 1) homologous expansion. 2) expansion opacities with optical depths using the Sobolev approximation. 3) Local thermodynamic equilibrium (LTE) is assumed for calculating the ionization and excitation states. The atomic line data for the bound-bound transitions, which constitutes the main and most important opacities for these problems, are taken from \citep[][CD 23]{kurucz95}. 
Using the extended line list of \citep[][CD 1]{kurucz94} has little effect on the optical light curve features investigated here (as noted by \citealt{k06}, the effect on the IR light curve is much more significant).
More details about the code, including convergence tests and comparisons to previously published radiative transfer codes for several benchmark problems are presented in appendix~\ref{sec:appendixA}. The program is publicly available and can be downloaded from \\
https://www.dropbox.com/sh/kyg1z1xwi0298ru/ AAAqzUMbr6AkoVfkSVIYChTLa?dl=0.

Light curves are calculated for a large set of ejecta including profiles from previous hydrodynamic calculations as well as synthetic ejecta where various physical parameters are scanned. The hydrodynamic model ejecta include series computed from the Chandrasekhar delayed-detonation calculations with varying deflagration to detonation transition densities of \cite{dbhk14} and the sub-Chandrasekhar, core detonation of white dwarfs with varying masses by \cite{s+10} (see table~\ref{table:ejecta_table_literature} for a summary of the main physical properties of these ejecta). 

We preformed radiation transfer calculations for a wide range of synthetic ejecta as described below and summarized in table \ref{table:ejecta_table} in appendix~\ref{sec:appendixB}.
Most of the synthetic ejecta are chosen based on 3 basic configurations which are modified in various aspects. 
The 3 nominal ejecta have \nickel masses: $M_{\rm Ni}=0.7M_{\odot}$, $M_{\rm Ni}=0.3M_{\odot}$ and $M_{\rm Ni}=0.1M_{\odot}$,  an exponential density profile $\rho\propto \exp(-v/v_e)$ where $v_e=(E_K/6M_{\rm ej})^{1/2}$ (e.g. \cite{wkbs07}) and a kinetic energy of $E_K=1.5\cdot10^{51}$ ergs. The nominal ejecta with \nickel masses $0.7$ and $0.3M_{\odot}$ (but not the $M_{\rm Ni}=0.1M_{\odot}$) have a central stable iron core of $0.1M_{\odot}$. The total ejecta masses ($M_{\rm ej}=1.21,1.08$ and $0.9 M_{\odot}$ respectively) were chosen such that the gamma-ray escape time is $t_0=35$ days. Outside the \nickel shell, the nominal ejecta were comprised of pure carbon (see section \S~\ref{sec:outer_layers_sensitivity}), with no mixing present between layers. Each of these configuration was then modified in various ways: 
\begin{enumerate}
	\item Changing the outer layer composition by either replacing the inner $0.3 M_{\odot}$ of carbon with a composition of IMEs (by mass: $0.53$ Si, $0.32$ S, $0.083$ Ca and $0.062$ Ar, as in \cite{wkbs07}) or by replacing the carbon with an equal mixture of oxygen and carbon. 
	
	\item Modifying the kinetic energy between $0.5\cdot10^{51}$ and $2\cdot10^{51}$ ergs and accordingly the mass so as to keep the gamma-ray escape time $t_0$ constant.
	\item Varying $t_0$ from 25 to 45 days by modifying the total mass.
	\item Changing the density profile to a constant density, also modifying the kinetic energy in the same range, while keeping $t_0$ constant. 
	\item Removing the central core of stable iron for the high \nickel mass configurations, or adding such a core of $M_{\rm Ni}=0.1M_{\odot}$ of stable iron for the case of low \nickel.
	\item For the configuration with only \nickel and Carbon (i.e. no central stable iron), mixing the \nickel and Carbon layers by different amounts (between 20 and 100 iterations of a moving box average, as described in \cite{wkbs07}).
\end{enumerate}

Following \cite{kw07}, we also simulated additional three series of ejecta in each of which the $^{56}$Ni abundance was multiplied by different constant factors (with lighter elements filling in) without otherwise affecting the structure or profile (these are termed 'Kasen-like' in the figures below).  All of these series have a range of total $^{56}$Ni of $0.1-0.7M_{\odot}$, kinetic energy of $10^{51}$ erg and gamma-ray escape time of $t_0=35$ days. Note that multiplying the $^{56}$Ni abundance by a constant factor does not change $t_0$ by construction.

The three series of this type are:
\begin{enumerate}
	\item An inner layer of $M_{\rm Ni}=0.1M_{\odot}$ stable iron, then a shell of \nickel and an outer pure Carbon layer, and for which the varying \nickel is replaced by IMEs (same composition as above and \citealt{wkbs07}). The total mass of these ejecta is $1.21M_{\odot}$ to achieve $t_0=35$ days .
	\item Same as the first series, but the varying \nickel is replaced by Carbon instead of IMEs. Same total mass of $1.21M_{\odot}$.
	\item Same as the first series but without the iron core - Inner core of \nickel surrounded by Carbon, varying $^{56}$Ni by uniformly replacing it with IMEs. The total mass of these ejecta is $1.13M_{\odot}$.
\end{enumerate}

\begin{figure*}
	\centering
	\includegraphics[width=\columnwidth]{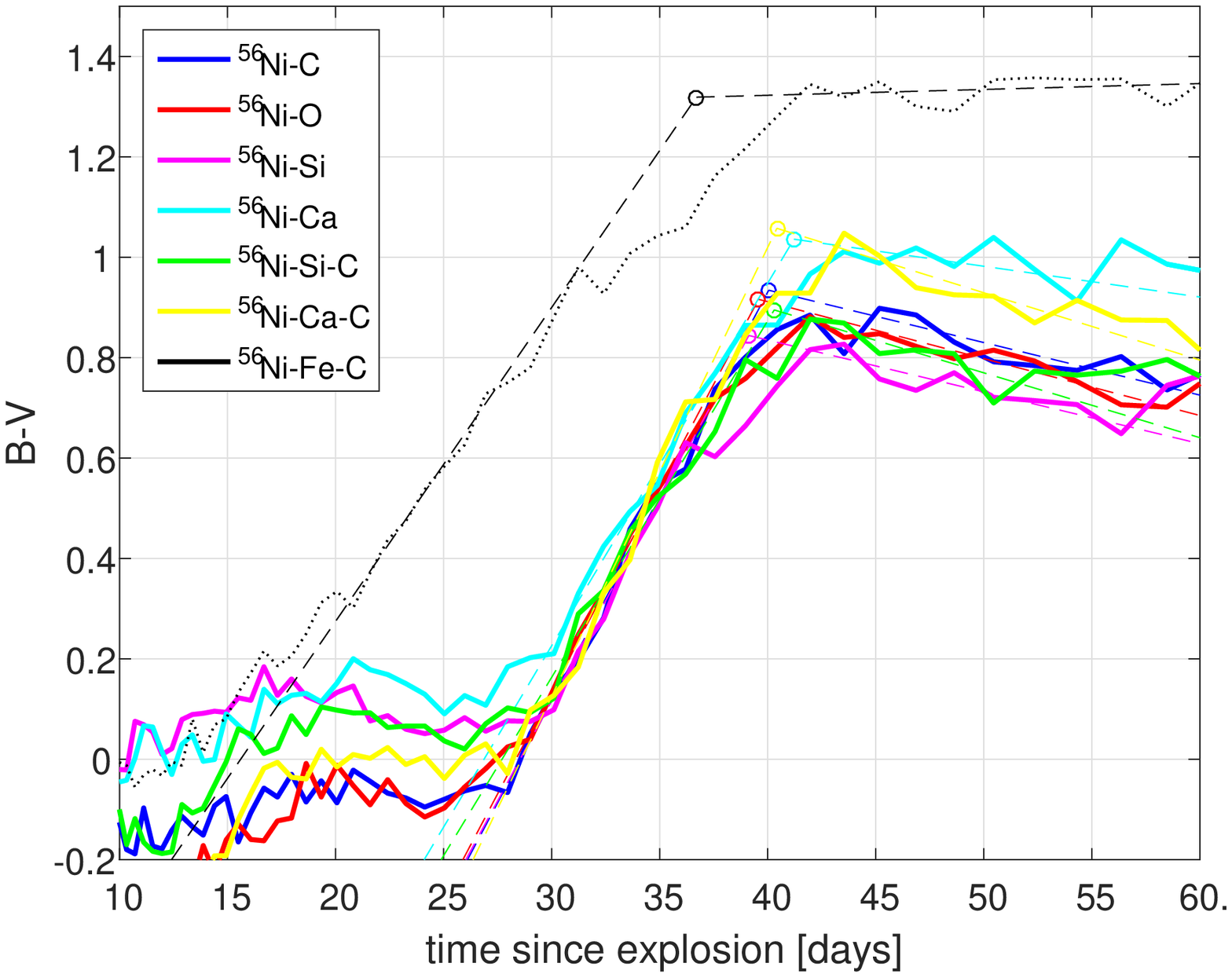}
	\hspace{2mm}
	\includegraphics[width=\columnwidth]{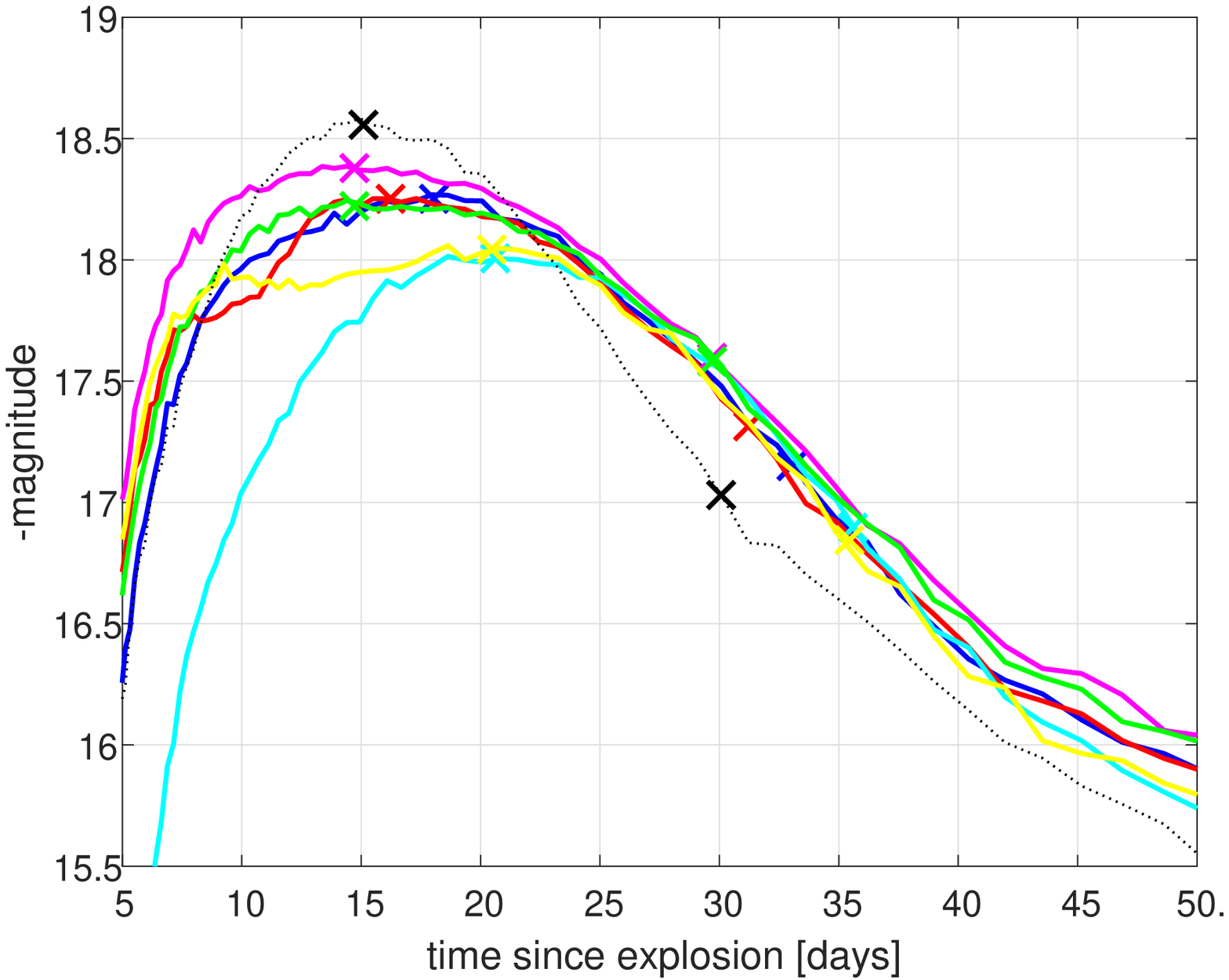}	
	\caption{left panel: B-V as a function of time for synthetic ejecta composed of $0.3M_{\odot}$ of \nickel surrounded by various compositions of the outer layers: $\sim0.7M_{\odot}$ of carbon (blue), oxygen (red), silicon (magenta) or calcium (cyan), or surrounded by a first layer of $0.3M_{\odot}$ of silicon (green), calcium (yellow) or iron (black) together with a second layer of carbon. All ejecta have the same density profile. Right panel: the B-band light curve for these same simulations, where for each curve \dmft was estimated and is indicated through crosses at the interpolated maximum of the curve and at its value after 15 days.}
	\label{fig:compare_compositions_BmV}
\end{figure*}

\subsection{sensitivity to outer layers composition}\label{sec:outer_layers_sensitivity}

in figure~\ref{fig:compare_compositions_BmV}, color (B-V, upper panel) and absolute B light curves (lower panel) are shown for ejecta with various combinations of C, O, Si, and Ca outside a core of $0.3M_{\odot}$ of \nickel (with $E_K=1.5\cdot10^{51}$ ergs and $t_0=$35 days). The small variance between these light curves, in particular around the time of maximum luminosity, causes differences of up to $\sim0.5$ mag in \dmft, which is significant compared to the total observed range of $\sim1$ mag. 
On the other hand, the B-V curve is much less dependent on the composition of the outer layers, and in particular the time indicator tbv varies by no more than 1 day. Significant amounts of iron at the outer parts have a much larger effect.  The results for an ejecta where a layer of $0.3M_{\odot}$ of stable iron surrounds the nickel core (likely unrealistic) are shown in dashed-black. Note that even in this extreme case, tbv is modified only by $\sim5$ days. 

The weak effect of the outer layers on tbv is related to the fact that at that epoch, the vast majority of the emitted photons had their last interaction in the Ni core. For the basic configuration shown here ($0.3M_{\odot}$ of \nickel surrounded by $\sim0.7M_{\odot}$ of carbon) only $13\%$ of the observed radiation has an interaction in the outer shell at 40 days. Around peak about $40\%$ of the photons have an interaction in the outer shell and therefore the light-curves are much more sensitive to the composition at those times.

We conclude that the properties of the light curve, and in particular tbv, are set mainly by the nickel amount and distribution in the ejecta.

\subsection{break time in B-V vs. recombination time}\label{sec:tbv_vs_recombination}
It has been shown qualitatively \citep{kw07, h+17} that the dynamics of the B-V curve (as well as the secondary infra-red peak, \citealt{k06})  is governed by the recombination of the iron group elements from double ionized to singly ionized states. We next revisit the relation between the recombination and the B-V curve focusing on the physical origin of the sharp break and its time. 

\begin{figure}
	\includegraphics[width=\columnwidth]{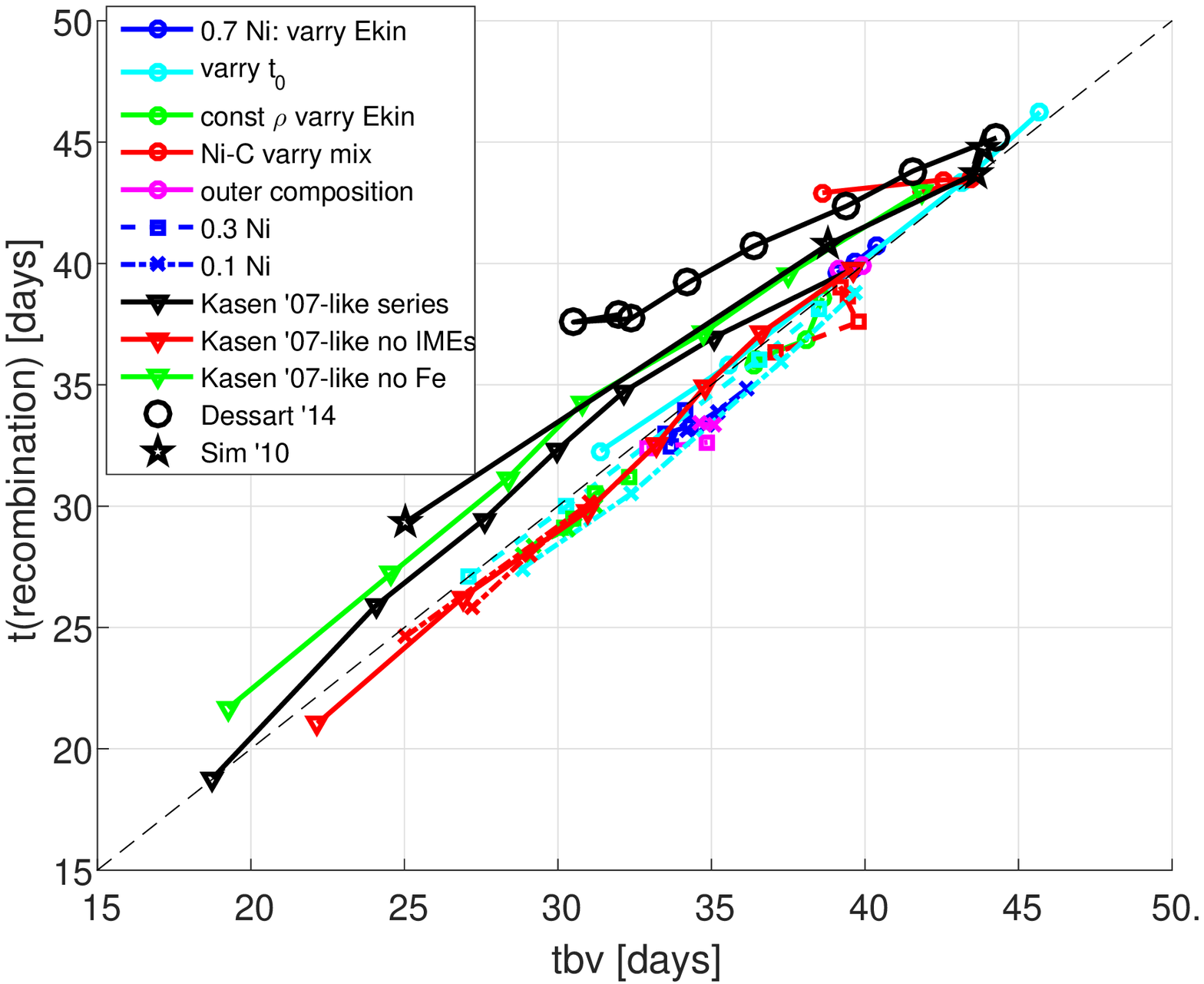}\llap{\raisebox{0.9cm}{\includegraphics[scale=0.2]{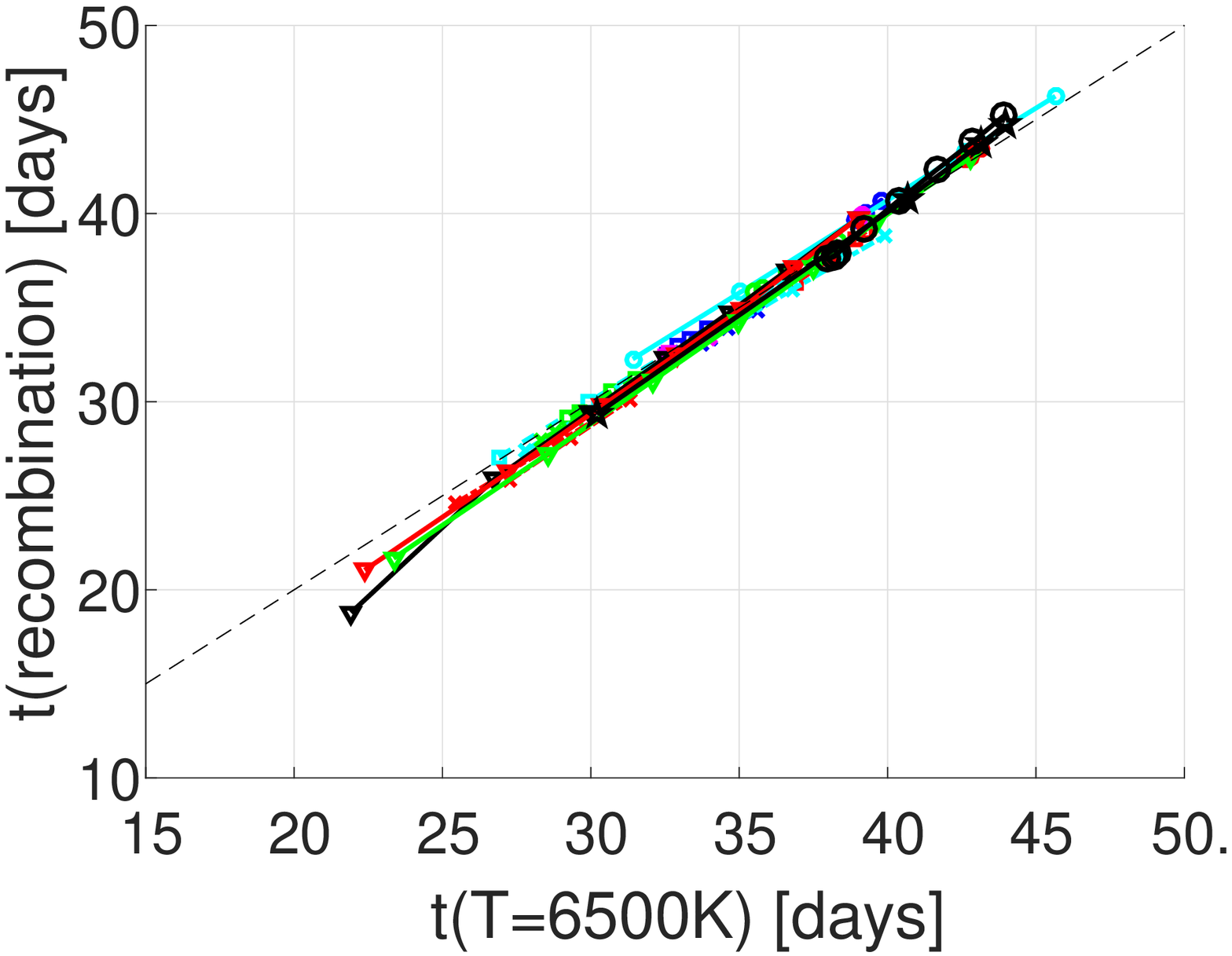}}}
	\caption{
		Mean recombination time of the \nickel in the ejecta compared to the B-V break time, tbv, for the simulated light curves described in \ref{sec:radiative_transfer}. Shown are ejecta from DD models \protect\citep{dbhk14} (black circles), from central detonations of sub-$M_{\rm ch}$ WDs \protect\citep{s+10} (black stars), as well as the large set of synthetic ejecta listed in table~\ref{table:ejecta_table}. The thin black dashed line represents the y=x line. \textbf{Inset:} Mean recombination time of the \nickel in the ejecta compared to the time when its mean temperature reaches $6500^oK$ for the same set of simulations.}
	\label{fig:tion_vs_tbv}
\end{figure}

In figure~\ref{fig:tion_vs_tbv}, a quantitative comparison is made between the B-V break time tbv and the time of recombination of \nickel and its decay products $^{56}$Co and $^{56}$Fe (henceforth collectively referred to as \nickel) in each of the simulated ejecta. The recombination time is defined here as the time at which the mass-weighted average ionization state of the \nickel is equal to 1.5. Note that adding stable iron-group elements to the weighted ionization has a negligible effect on the recombination time calculation and they are not included. As can be seen in the figure, tbv and the time of recombination are approximately equal in all calculations and differ by no more than 5 days for all cases except for the low \nickel $M_{\rm ch}$ DD ejecta from \cite{dbhk14}. In these configurations, the ejecta contain a large amount of stable iron mixed with the \nickel throughout much of the ejecta, and there are significant deviations  $\sim10$ days between the B-V break break time tbv and the recombination time of \nickel. The presence of stable iron up to high velocities modifies the light curve color significantly, as is shown to be the case in a synthetic example in figure~\ref{fig:compare_compositions_BmV}. Note that when stable iron is present but concentrated in the inner part of the ejecta, its effect on the light curve is negligible.

The recombination reflects a global change in the ejecta and occurs when the mean temperature of the \nickel reaches $T\approx6500K$. This is demonstrated in the inset in figure~\ref{fig:tion_vs_tbv} where the recombination time is shown vs the time at which the mass-weighted average temperature of \nickel and its decay products reaches $6500K$. Evidently, the variations by factors of a few in the densities between the ejecta have little effect on the recombination temperature. 

\begin{figure*}
	\includegraphics[width=\columnwidth]{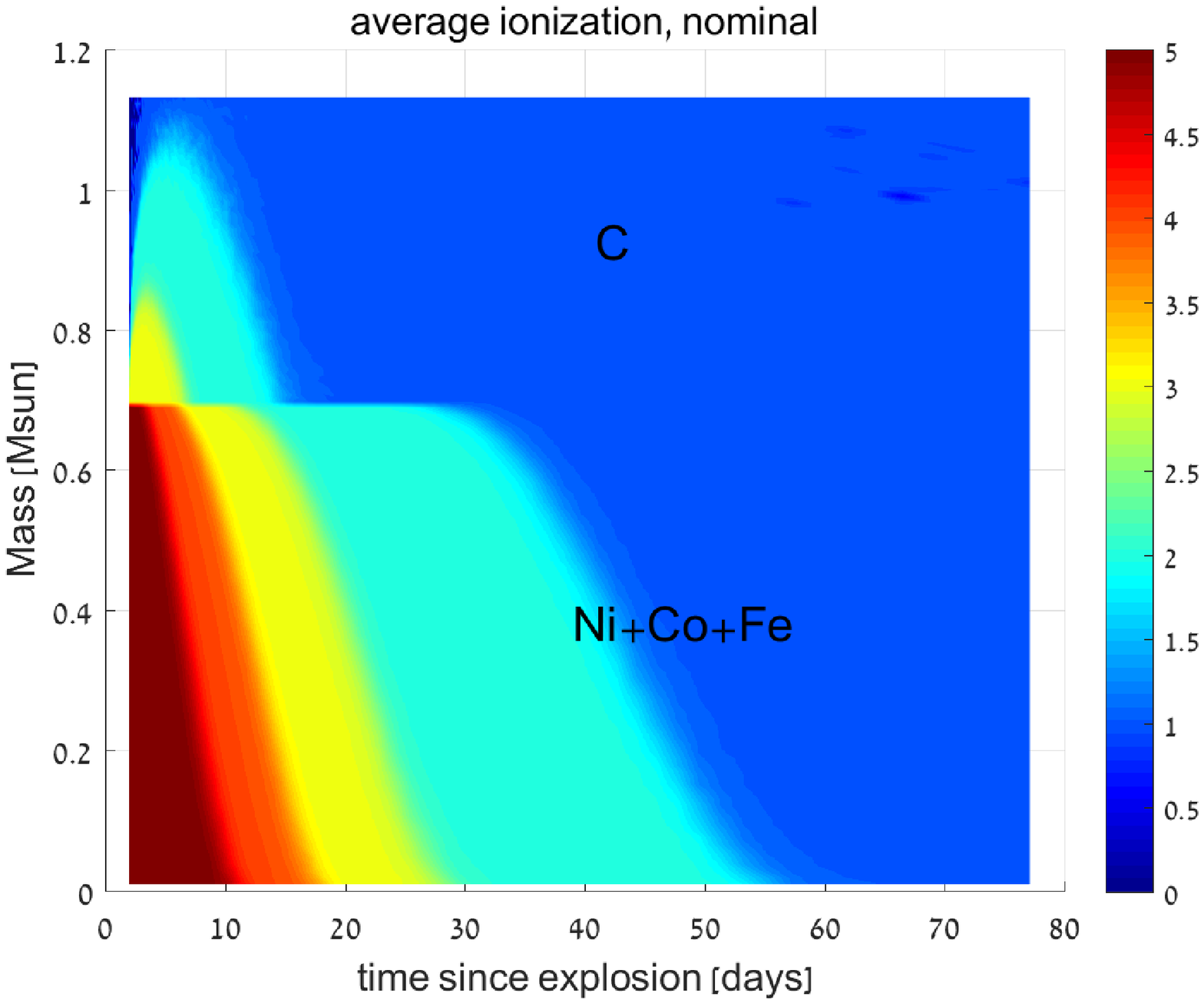}
	\hspace{2mm}
	\includegraphics[width=\columnwidth]{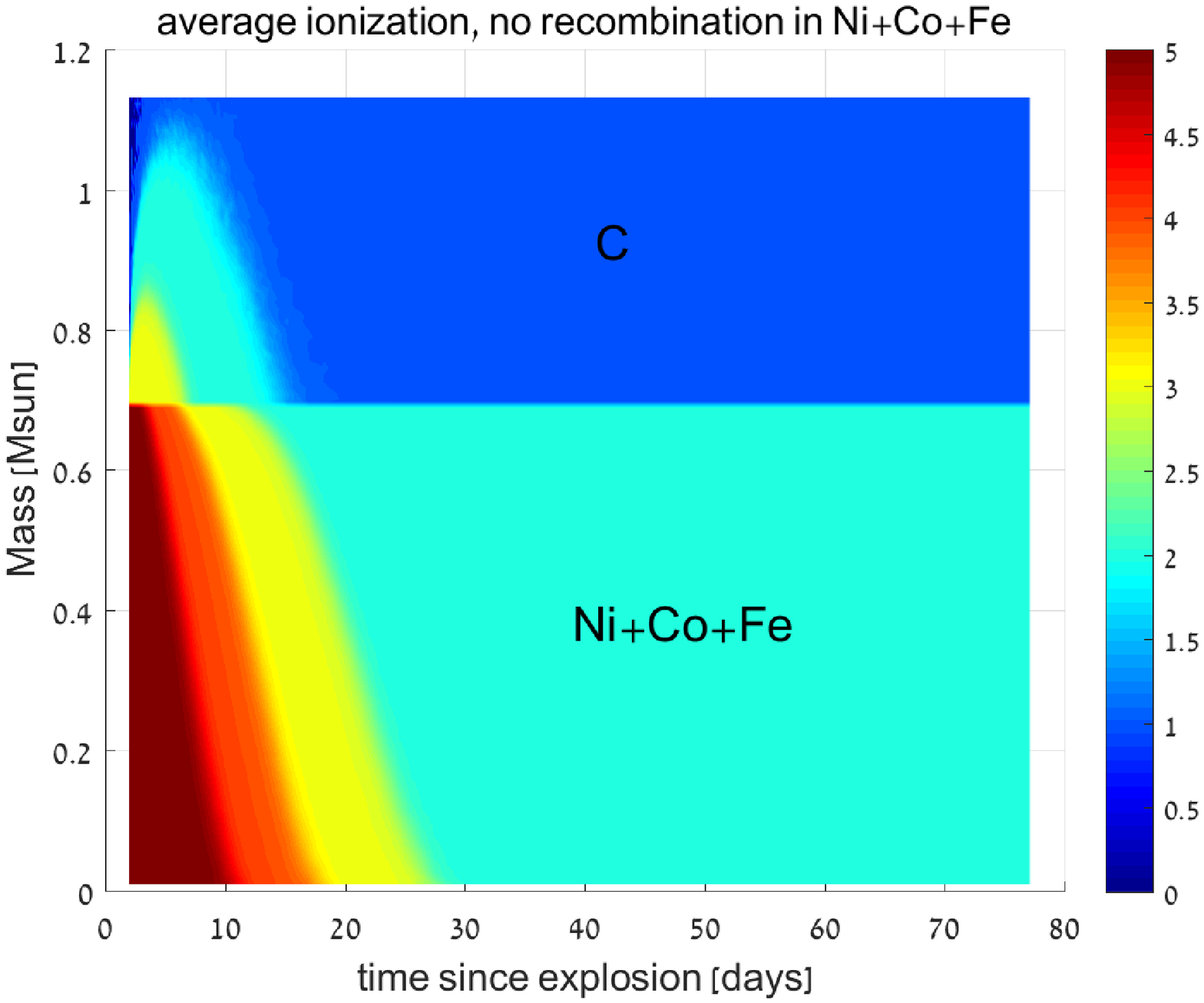}
	\vspace{2mm}	
	\includegraphics[width=\columnwidth]{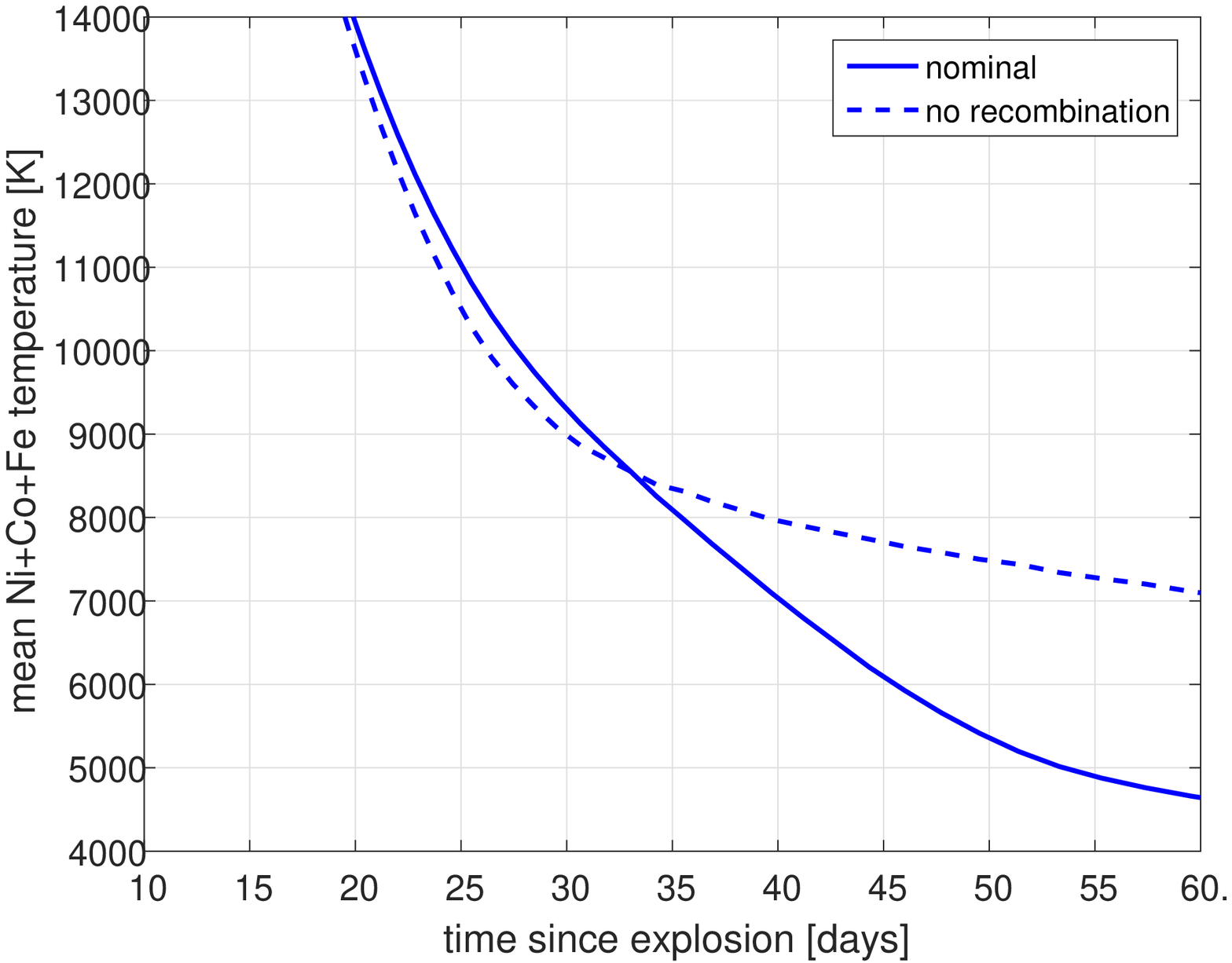}
	\hspace{2mm}
	\includegraphics[width=\columnwidth]{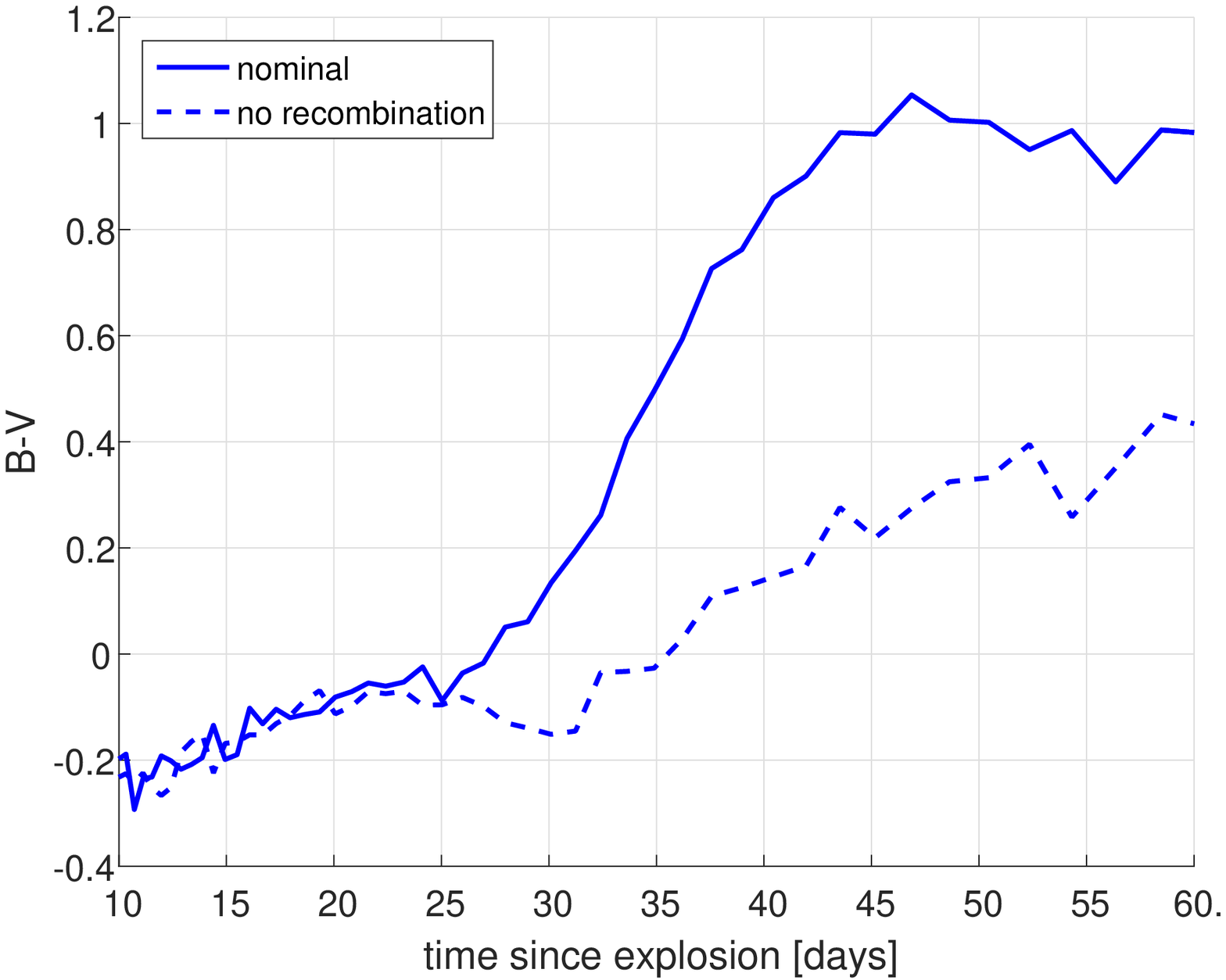}
	\vspace{2mm}
	\includegraphics[width=\columnwidth]{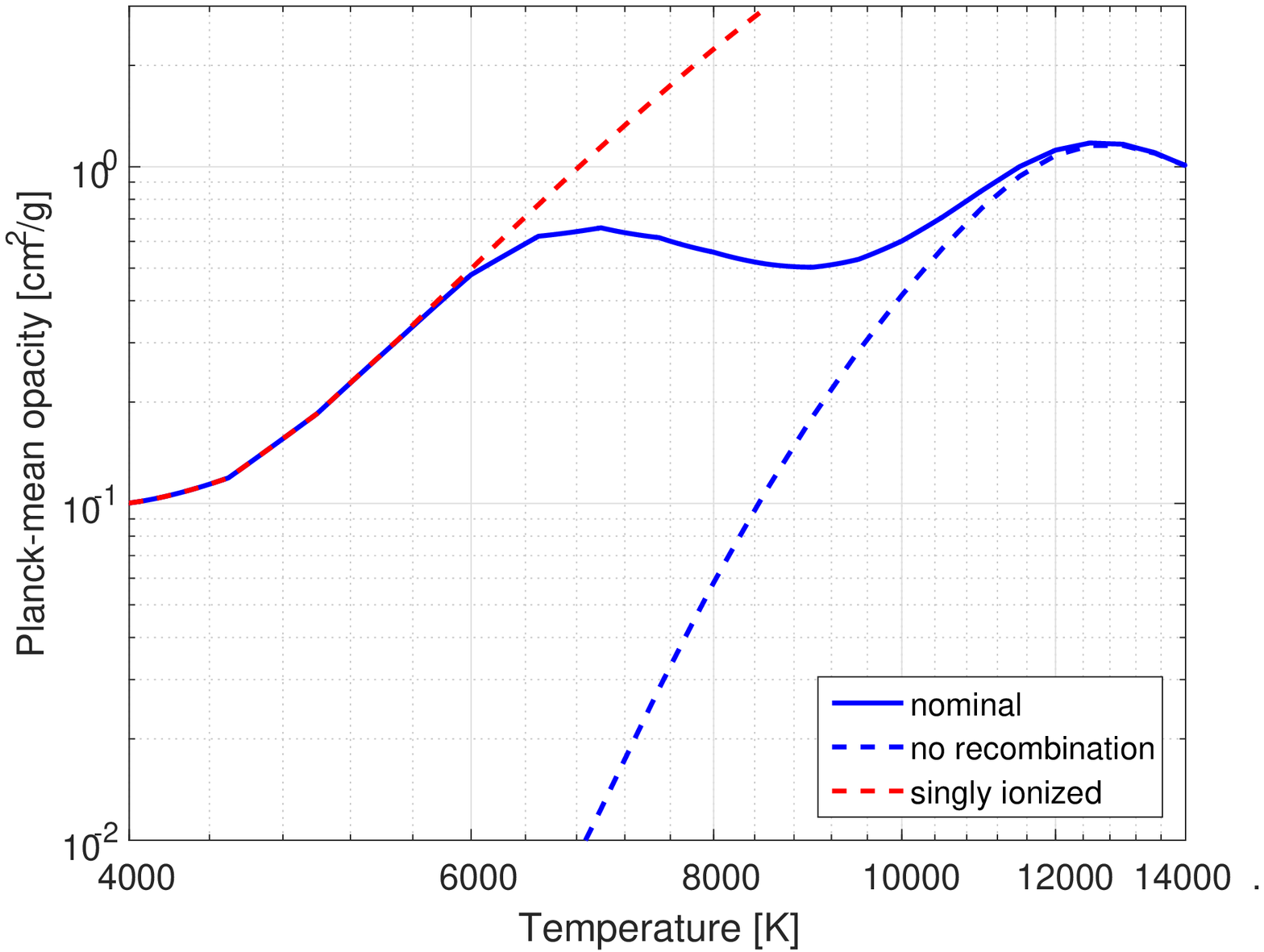}
	\hspace{2mm}
	\includegraphics[width=\columnwidth]{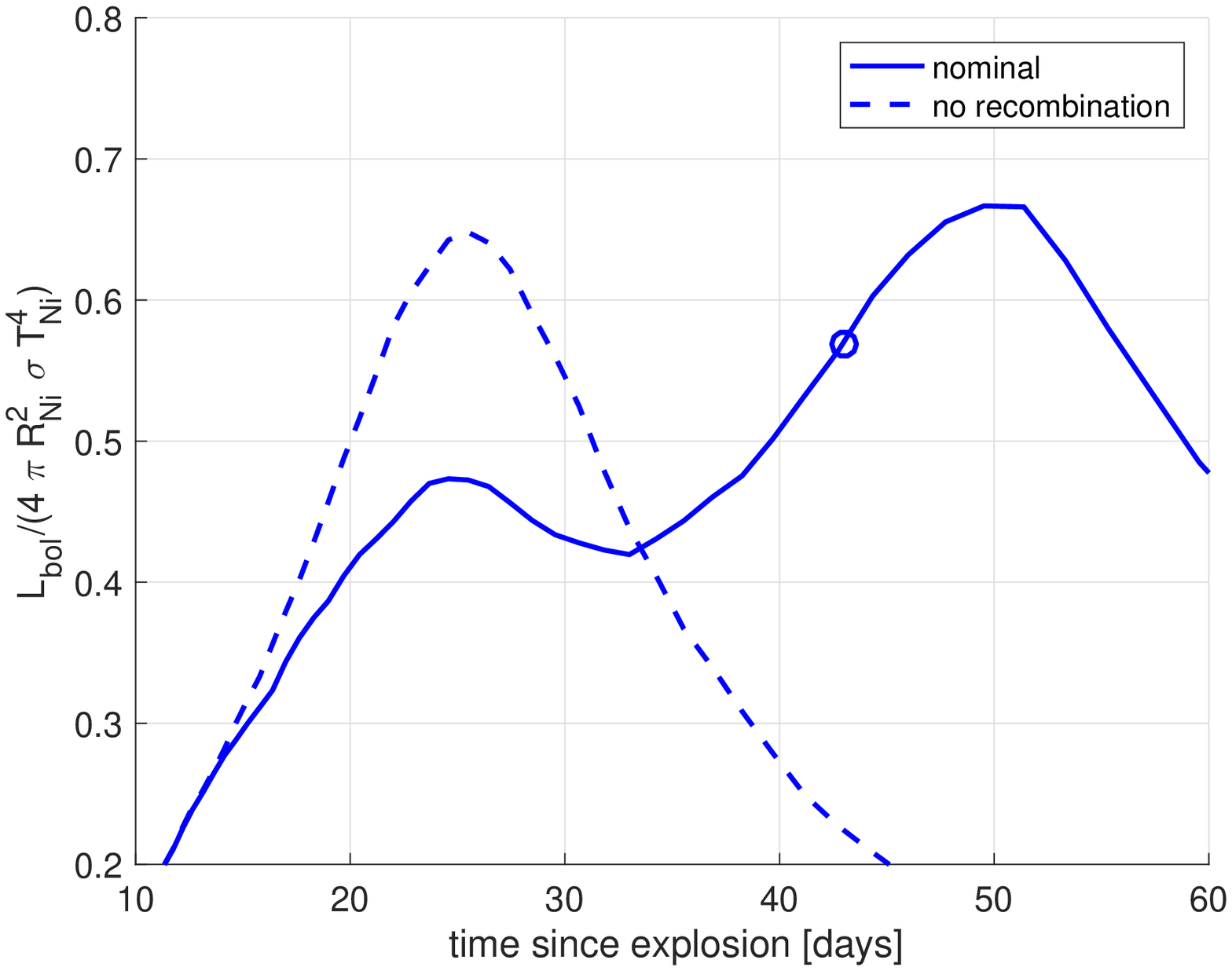}
	\caption{The physical effects of recombination are studied by performing radiation transfer calculations for a synthetic ejecta with \nickel=0.7$M_{\odot}$ and $t_0=35$ days where the recombination of iron group elements is artificially prevented. The results are compared to a standard (nominal) calculation of the same ejecta.
		\textbf{top row:} color maps of the ionization state as a function of time (X-axis) and mass (Y-axis) for the standard calculation (left) and with recombination prevented (right). 
		\textbf{center-left} Mean temperature of the iron-group elements in the two simulations. \textbf{center-right:} B-V light curves as a function of time. \textbf{bottom left:} mean planck opacity as function of temperature assuming a composition of the decay products of $^{56}$Ni at 35 days and density of $\rho=10^{-14}\rm g~cm^{-3}$. \textbf{bottom right:} Ratio of bolometric luminosity $L_{\rm bol}$ and averaged black-body photospheric estimate using the ($^{56}$Ni weighted) average temperature $T_{\rm Ni}$ and radius $R_{\rm Ni}$. The recombination time is marked by an open circle. }
	\label{fig:no_recombination_effect}
\end{figure*}

To further illustrate the effect of recombination on the color curve, a simulation was preformed where the recombination to the singly ionized state was prevented by artificially lowering the second ionization energies (setting them to be identical to the first ionization energies). This calculation was preformed for a synthetic ejecta with $0.7M_{\odot}$ of \nickel surrounded by carbon with $t_0=35$ days (line 6 in table~\ref{table:ejecta_table}). The results are compared to a standard (nominal) calculation of the same ejecta, where recombination was allowed, in figure \ref{fig:no_recombination_effect}. The top color panels show the ionization dynamics as a function of time and mass coordinate for the standard calculation (upper left panel) and the no-recombination calculation (upper right panel). Comparing the mean \nickel temperature (middle-left panel) illustrates clearly that starting at $\sim35$ days from explosion, the temperature decline is faster in the standard calculation where recombination occurs.  

The lower-left panel presents the Planck-weighted average opacities for \nickel  as a function of temperature (with decay products taken at 35 days, and density set to $\rho=10^{-14}$ gr/cc). As can be seen, the opacity of both the pure singly (dashed red-line) and doubly (case of no recombination, blue-dashed) ionization states rapidly drops at temperatures $T\lesssim 10000$K with the singly ionized elements having much higher opacities. In the full calculation (i.e. ionization state is temperature dependent as well), the rapid rise of singly ionized elements as the temperature decreases in the range $10000-7000$K, compensates for the strong temperature dependence of the opacity in each ion and results in a roughly constant opacity in this range. Once the temperature reaches $T\approx 6500$K, a significant fraction of the mass is in singly ionized state leaving no room to further rapid increase in the singly ionized fraction. A cliff is obtained with lower temperatures having much smaller opacities. In the calculation with no recombination there is no compensation and the opacity strongly depends on the temperature at all temperatures $T\lesssim 10000$K. 

The center-right panel compares the resulting B-V curves for the two simulations. The  B-V curve in the standard calculation features a phase of $\sim20$ days of sharp reddening (i.e. B-V rising), corresponding to the temperature range where the opacity is roughly constant, followed by a phase of relatively constant color. The break in the nominal B-V curve occurs at the time when the mean temperature is $\sim6500^oK$, which is the temperature at which the opacity cliff is present. In the case where recombination is inhibited, the opacities decline with temperature monotonically and the rise is much more gradual- no noticeable break is present in the curve.

The bottom right panel shows the evolution of the ratio of total luminosity $L_{\rm bol}$ to an averaged black-body photospheric estimate using the ($^{56}$Ni weighted) average temperature $T_{\rm Ni}$ and radius $R_{\rm Ni}$, $L_{\rm BB}=4\pi R_{\rm Ni}^2\sigma T_{\rm Ni}^4$.  
At early times, when the ejecta is optically thick, the photosphere is close to its outer edge, where the temperature is lower than the mean \nickel temperature. Thus, estimating the photospheric temperature as the mean temperature, as done here, overestimates the luminosity, and $L_{\rm BB}\gg L_{\rm bol}$ (more than the underestimation of taking the radius as the mean radius rather than the photospheric one). At late times, when the ejecta is optically thin, the black body over estimates the emission and again $L_{\rm BB}\gg L_{\rm bol}$. Around the transition from the optically thick to optically thin regimes (see also \cite{h+17}), this ratio reaches a maximum value of order unity, at a time which coincides with the recombination time (marked by an open circle in the figure). As can be seen, the calculation with no recombination, where the opacities are much lower, transitions to the optically thin regime much earlier than the standard calculation.

\begin{figure}
	\includegraphics[width=\columnwidth]{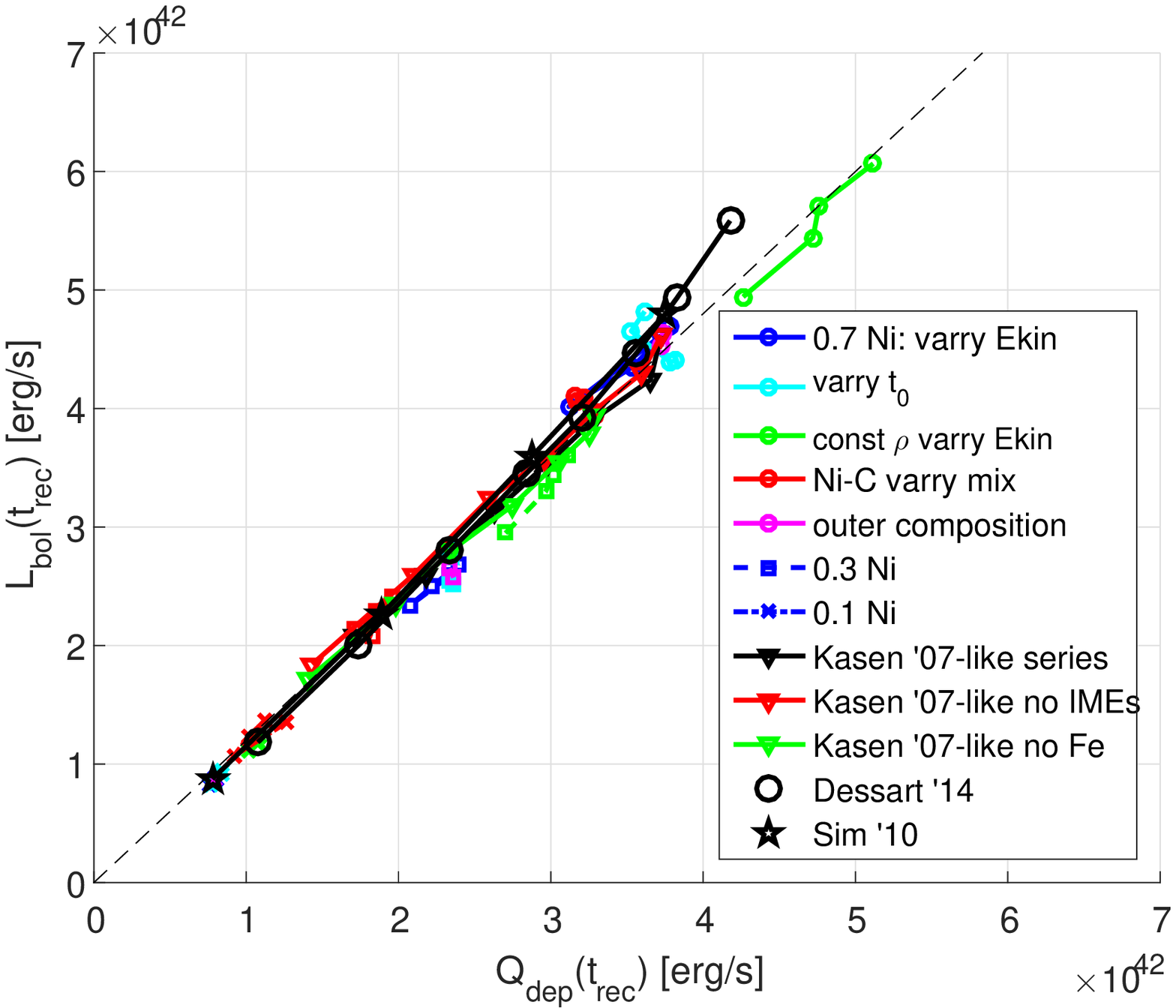}
	\caption{Bolometric luminosity at time of recombination versus energy deposition rate at this time. the dashed black line represents $y=1.2\cdot x$.}
	\label{fig:Lion_over_Qdep_vs_tion}
\end{figure}

A qualitative physical picture for the shape of the B-V curve can be described as follows. First note that around the recombination time, the bolometric luminosity is approximately equal to the energy deposition rate (see figure \ref{fig:Lion_over_Qdep_vs_tion}) and is slowly monotonically decreasing. 
At early times, the $^{56}$Ni core is optically thick. The decreasing luminosity requires a decreasing surface temperature which (very) roughly follows the black-body relation $T\propto L^{1/4}$. In the calculation in which recombination is prohibited, the rapidly declining opacities cause the ejecta to become optically thin at an early phase around $25$ days (see dashed line in the lower-right panel). Further temperature decline is stalled since the emissivity (proportional to the opacity) needs to maintain the slowly declining luminosity. In the realistic calculation where recombination is allowed, the prolonged higher opacities allow the temperature to continue to decline in a longer phase of large optical depth. The color becomes redder with time due to the declining temperature. The reddening rate is even faster than expected by a black body spectrum, since the increasing emissivities from the hotter center of the $^{56}$Ni manage to leak out at longer wavelengths as the ejecta is becoming less and less optically thick. The temperature decline is finally stalled once the opacity reaches the cliff around $T=6500$K. The transition from declining temperature to a stalling temperature is what causes the break in the B-V curve. 

\section{time of recombination can be simply predicted}\label{sec:tbv_predicted}

In this section we derive a simple analytical model that allows the recombination time to be estimated from the properties of the ejecta without the need for radiation transfer. 

\subsection{bolometric luminosity}
As can be seen in figure \ref{fig:Lion_over_Qdep_vs_tion} the bolometric luminosity at recombination time is to a good approximation equal to the instantaneous energy deposition rate from gamma-rays and positrons $Q_{\rm dep}$ at the time of recombination (note a systematic offset of about 20\%),\begin{equation}
L_{\rm bol}(t_{\rm rec})\approx Q_{\rm dep}(t_{\rm rec}).
\label{eq:LeqQ}
\end{equation}
The deposition rate is set by two parameters, the \nickel mass and $t_0$ \citep[e.g.][]{j99}, and can be approximated to an accuracy of better than $\pm10\%$ at all times (see paper I) using the interpolation:
\begin{equation}
Q_{\rm dep}=M_{\rm Ni56}\left(q_{\gamma}(t)\cdot(1-e^{-\frac{t_0^2}{t^2}})+q_{\rm pos}(t)\right),
\label{eq:qdep}
\end{equation}
where $q_{\gamma}$ and $q_{\rm pos}$ are the (normalized) total energy release rate of gamma-rays and positron kinetic energy respectively and are given by:
\begin{align}
q_{\gamma}(t)&=\left[6.45~e^{-\frac{t}{8.76\rm d}}+1.38~e^{-\frac{t}{111.4\rm d}}\right]\times 10^{43}\rm erg~s^{-1}~M_{\odot}^{-1},\cr
q_{\rm pos}(t)&=4.64\left[-e^{-\frac{t}{8.76\rm d}}+e^{-\frac{t}{111.4\rm d}}\right] \times10^{41}\rm erg~s^{-1}~M_{\odot}^{-1}.\cr
\label{eq:qgamma}
\end{align}

\subsection{photospheric emission}
Given that at recombination, the ejecta is marginally optically thick (see \S~\ref{sec:tbv_is_recombination}) the luminosity can be roughly approximated as a black-body emission from a photosphere
\begin{equation}
L_{\rm bol}\sim 4\pi R_{\rm Ni} ^2\sigma_{\rm sb} T_{\rm Ni} ^4.
\label{eq:LbolLBB}
\end{equation}
which is located at the ($^{56}$Ni weighted) mean radius 
\begin{equation}
R_{\rm Ni}\equiv \big<r\big>_{^{56}\rm Ni}~=~ v_{\rm Ni}\times t
\end{equation}
with a temperature equal to the average temperature $T_{\rm Ni}$ of the nickel at the time of recombination,
\begin{equation}
T_{\rm Ni}\equiv \big<T\big>_{^{56}\rm Ni}=6500K.
\end{equation}

\begin{figure}
	\includegraphics[width=\columnwidth]{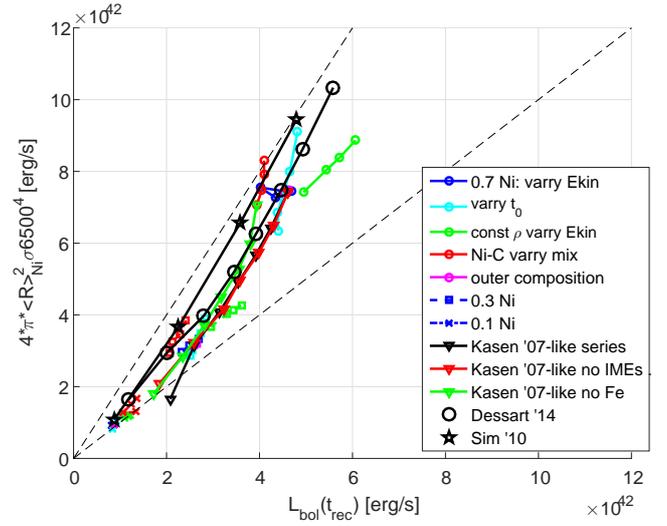}
	\caption{Ratio between the bolometric luminosity at recombination time and a simple photospheric emission assuming a radius equal to the mean \nickel radius $R_{\rm Ni}$ at that time and a temperature of 6500K, 
		for all simulated ejecta. The dashed black lines represent y=x and y=2x.}
	\label{fig:Ltbv_vs_photospheric}
\end{figure}

In figure~\ref{fig:Ltbv_vs_photospheric}, the ratio between the bolometric luminosity at the time of recombination and the estimate in equation \eqref{eq:LbolLBB} is shown for the whole set of simulated ejecta. It is evident in the figure that this ratio in the range of $\approx 0.5-1$ for all simulated ejecta. 

\subsection{inferring the recombination time}
The recombination time can be inferred by equating the energy deposition rate to the photospheric emission at recombination time (i.e. at 6500K):
\begin{equation}
Q_{\rm dep}(t)=f\cdot 4\pi (v_{\rm Ni} \cdot t)^2\sigma_{\rm sb}\cdot 6500K^4,
\label{eq:trec}
\end{equation}
where $f$ is a dimensionless number of order unity that accounts for the fact that: a) the bolometric luminosity is typically $\sim20\%$ higher than the deposition rate at recombination time, and b) the bolometric luminosity is lower than the rough photospheric estimate in Eq. \eqref{eq:LbolLBB} by a factor of 1-2. 

\begin{figure}
	\includegraphics[width=\columnwidth]{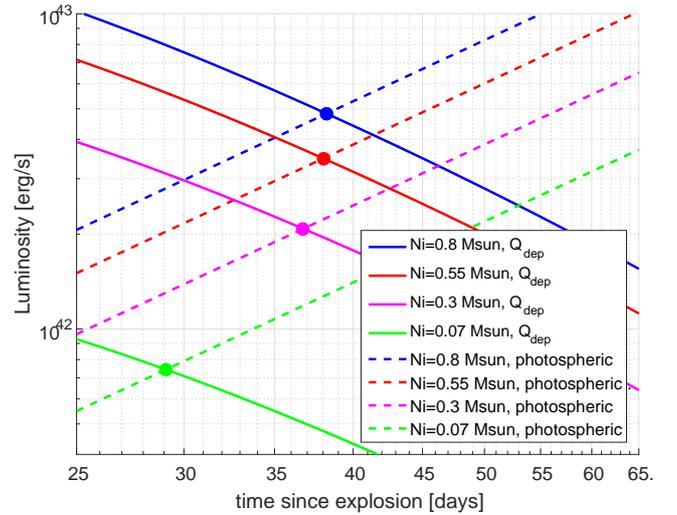}
	\caption{Visualizing the calculation of the recombination time for the four ejecta from \protect\cite{s+10} having \nickel masses of 0.8, 0.55, 0.3 and 0.07 solar masses (blue, red, magenta and green lines). For each ejecta, the rate of deposition of energy in the ejecta (\qdep, solid line) and the photospheric emission from the 'mean' iron group elements (dashed line) are shown. The recombination time is estimated as the time at which these two luminosities are equal (marked with a thick dot for each ejecta).}
	\label{fig:photospheric_visual}
\end{figure}

The deposition $Q_{\rm dep}(t)$ is declining in time and proportional to $M_{\rm Ni56}$ (and depends also on the gamma-ray escape time $t_0$). The photospheric emission is rising with time and proportional to $4\pi v_{\rm Ni}^2$. The recombination time is the time at which the two functions meet and depends on the ratio $M_{\rm Ni56}/4\pi v_{\rm Ni}^2$ as can be seen in figure \ref{fig:photospheric_visual} where $Q_{\rm dep}(t)$ and the photospheric emission are shown for the calculations of the sub-$M_{\rm ch}$ ejecta from  \cite{s+10}.

\begin{figure*}
	\centering
	\includegraphics[width=\columnwidth]{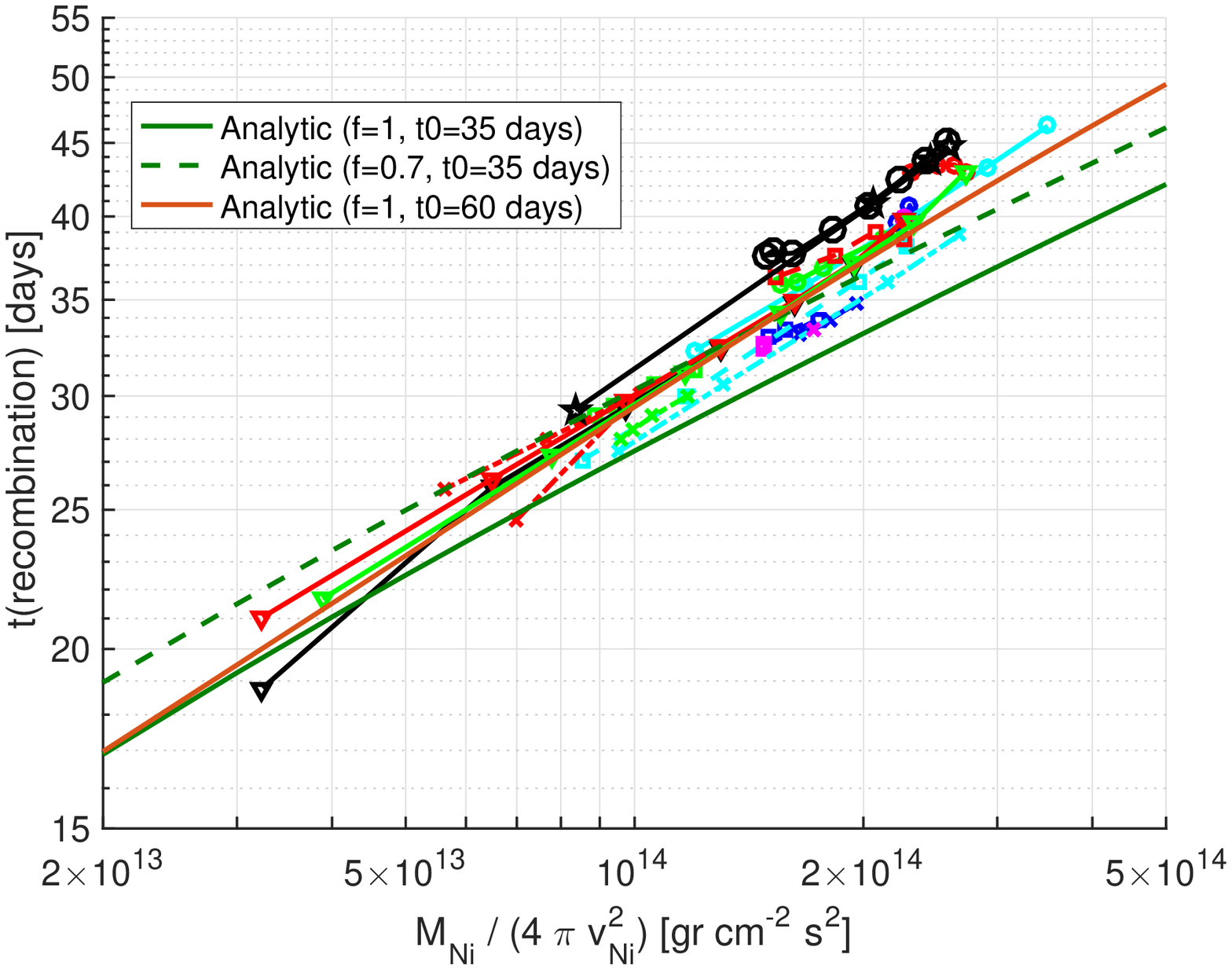}
	\hspace{2mm}
	\includegraphics[width=\columnwidth]{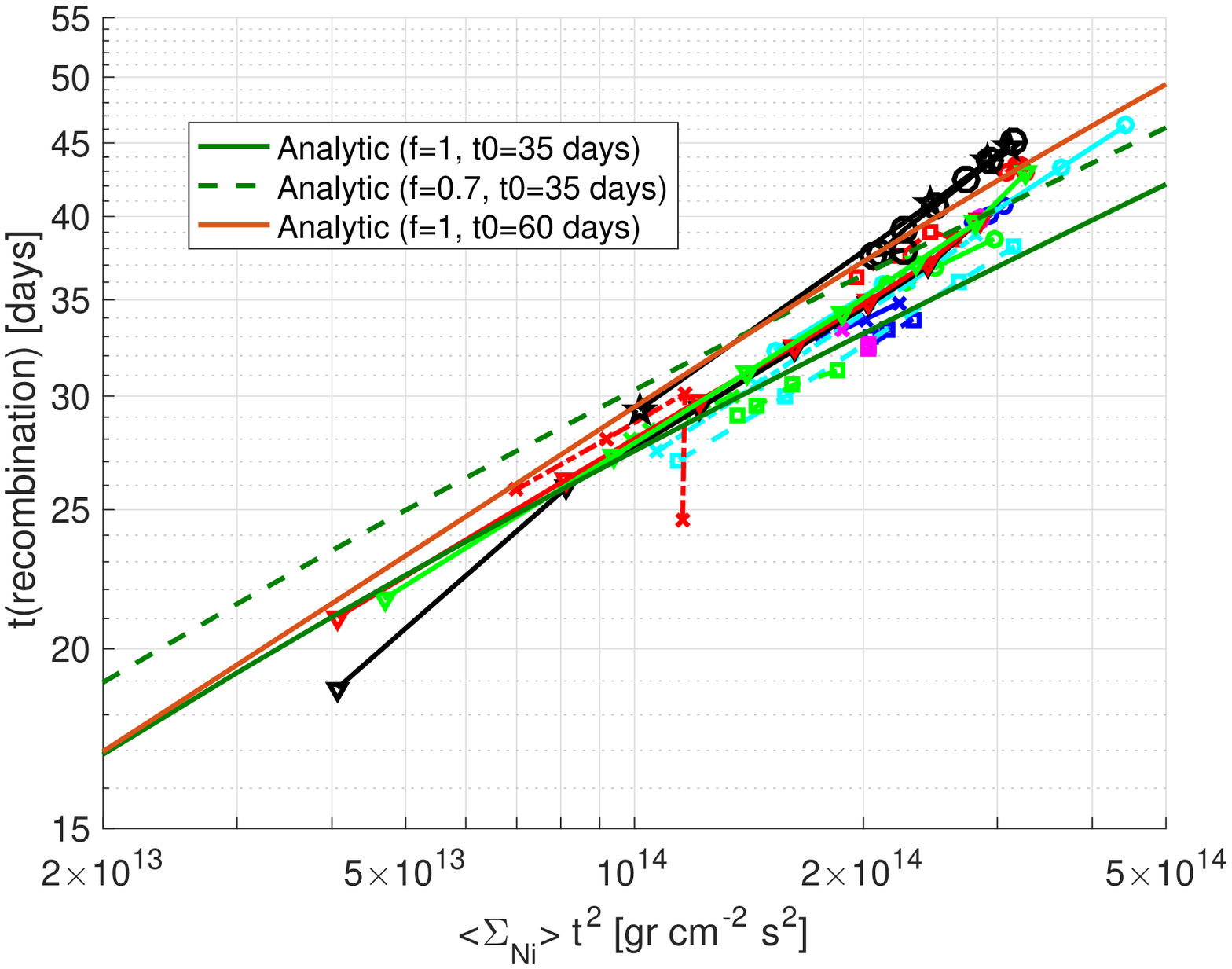}	
	\caption{
		The recombination time as a function of the nickel column density. The results of all simulated ejecta are shown for two parameterizations of the $^{56}$Ni column density: \textbf{left:} $M_{\rm Ni}/(4\pi \langle v \rangle_{\rm Ni}^2)$. \textbf{right:}  $\Sigma_{\rm Ni}\cdot t^2$ as calculated in Eq. \eqref{eq:SigmaNi}. The result of the simple photospheric emission model, Eq. \eqref{eq:trecequation}, is shown for three choices of parameters: Nominal calculation ($f=1$) with $t_0=35$days (solid green line), calculation with single fitting parameter $f=0.7$, also with $t_0=35$days (dashed green line), and nominal calculation ($f=1$) with $t_0=60$days (solid brown line).}
	\label{fig:trec_vs_sigmanini}
\end{figure*}

To isolate the dependence on the properties of the ejecta, it is useful to rewrite Eq. \eqref{eq:trec} using \eqref{eq:qdep}: 
\begin{equation}
q_{\gamma}(t)(1-e^{-\frac{t_0^2}{t^2}})+q_{\rm pos}(t)=f\left(\frac{M_{\rm Ni56}}{4\pi v_{\rm Ni}^2}\right)^{-1}\sigma_{\rm sb}~ 6500K^4~t^2
\label{eq:trecequation}
\end{equation}
where $q_{\gamma}$ and $q_{\rm pos}$ are functions of time given by Eq. \eqref{eq:qgamma}.  The results of equation \eqref{eq:trec} are compared to the recombination times obtained in the radiation transfer simulations in the left panel of figure \ref{fig:trec_vs_sigmanini} for fixed values of $t_0=35d,60d$ and $f=0.7,1$. Note that there is little sensitivity to the values of $t_0$ and $f$ and that setting $f=0.7$ is a reasonable approximation for the whole set of simulated ejecta. Since the deposition rate (l.h.s) declines strongly as a function of time, while the photospheric emission (r.h.s) rises strongly with time, the scatter in the inferred recombination times from equating these two is small despite the fact that for different ejecta, the ratio between these two was shown to vary by up to a factor of 2. 

As explained above and as can be seen from Eq. \eqref{eq:trecequation} or figure \ref{fig:trec_vs_sigmanini}, the recombination time is set by the combination $M_{\rm Ni56}/4\pi v_{\rm Ni}^2$, which is roughly the (undecayed) $^{56}$Ni column density (see clarifications below). \emph{The fact that the B-V break time is equal to the recombination time and is set by the $^{56}$Ni column density are the main results of this paper}.

Two points need to be clarified regarding the $^{56}$Ni column density. First, any column density $\Sigma\propto M/R^2$ in homologous expansion declines with time exactly as $t^{-2}$. Here we interchangeably use the term column density for the time independent expression $\Sigma \cdot t^2\propto M/v^2$. Second, the column density along a ray from some point in the ejecta depends on the position in the ejecta and the direction of the ray. The expression $M_{\rm Ni56}/4\pi v_{\rm Ni}^2$ is one choice for a rough average of the column densities of $^{56}$Ni across the ejecta. We next show that there is little sensitivity to the precise choice.

The gamma-ray escape time $t_0$, that sets the bolometric light-curve width, is also a function of a column density - the total column density $\sim M/4\pi v^2$. In fact, $t_0$ is precisely given by \cite{j99}
\begin{equation}\label{eq:t0Sigma}
t_0=\sqrt{\kappa_{\gamma}\langle \Sigma \rangle t^2} 
\end{equation}
where $\langle \Sigma \rangle\propto t^{-2}$ is the average column density of the ejecta as seen by the $^{56}$Ni elements
\begin{equation}\label{eq:SigmaNi}
\langle\Sigma\rangle=\int \frac{d^3x}{M_{\rm Ni56}}~\rho_{\rm Ni56}({\bf x})\int \frac{d{\bf\hat\Omega}}{4\pi} \int_0^{\infty} ds~\rho({\bf x}+s{\bf \hat \Omega})
\end{equation}
where $\rho_{\rm Ni56}$ is the (un-decayed) density of $^{56}$Ni and $\rho$ is the total density. The effective opacity $\kappa_{\gamma}=0.025 \rm cm^2/g$ is calculated by averaging the Klein-Nishina corrected Compton cross section and average fractional energy loss per scattering over each of the (discrete) emitted gamma-ray energies $E_n$ \citep{s+95,j99}. To better relate the two time scales (recombination and gamma-escape time) it is useful to define an equivalent $^{56}$Ni column density, $\langle \Sigma_{\rm Ni} \rangle$ in the same way:
\begin{equation}\label{eq:SigmaNiNi}
\langle\Sigma_{\rm Ni}\rangle=\int \frac{d^3x}{M_{\rm Ni56}}~\rho_{\rm Ni56}({\bf x})\int \frac{d{\bf\hat\Omega}}{4\pi} \int_0^{\infty} ds~\rho_{\rm Ni56}({\bf x}+s{\bf \hat \Omega})
\end{equation}
Both $\langle \Sigma\rangle$ and $\langle \Sigma_{\rm Ni} \rangle$ involve integration along the same rays with the difference being that along each ray,  $\langle \Sigma \rangle$ involves the integration of density and $\langle \Sigma_{\rm Ni} \rangle$ involves the integration of the $^{56}$Ni density. The nickel column density $\langle \Sigma_{\rm Ni} \rangle t^2$ is approximately equal ($0.8 \pm 0.2$ for all synthetic and explosion model ejecta, including two dimensional ejecta from WDs collisions)
to the estimate $M_{\rm Ni56}/4\pi v_{\rm Ni}^2$ and it makes little difference which one is used (see figure \ref{fig:trec_vs_sigmanini}). 

\begin{figure*}
	\centering
	\includegraphics[width=\columnwidth]{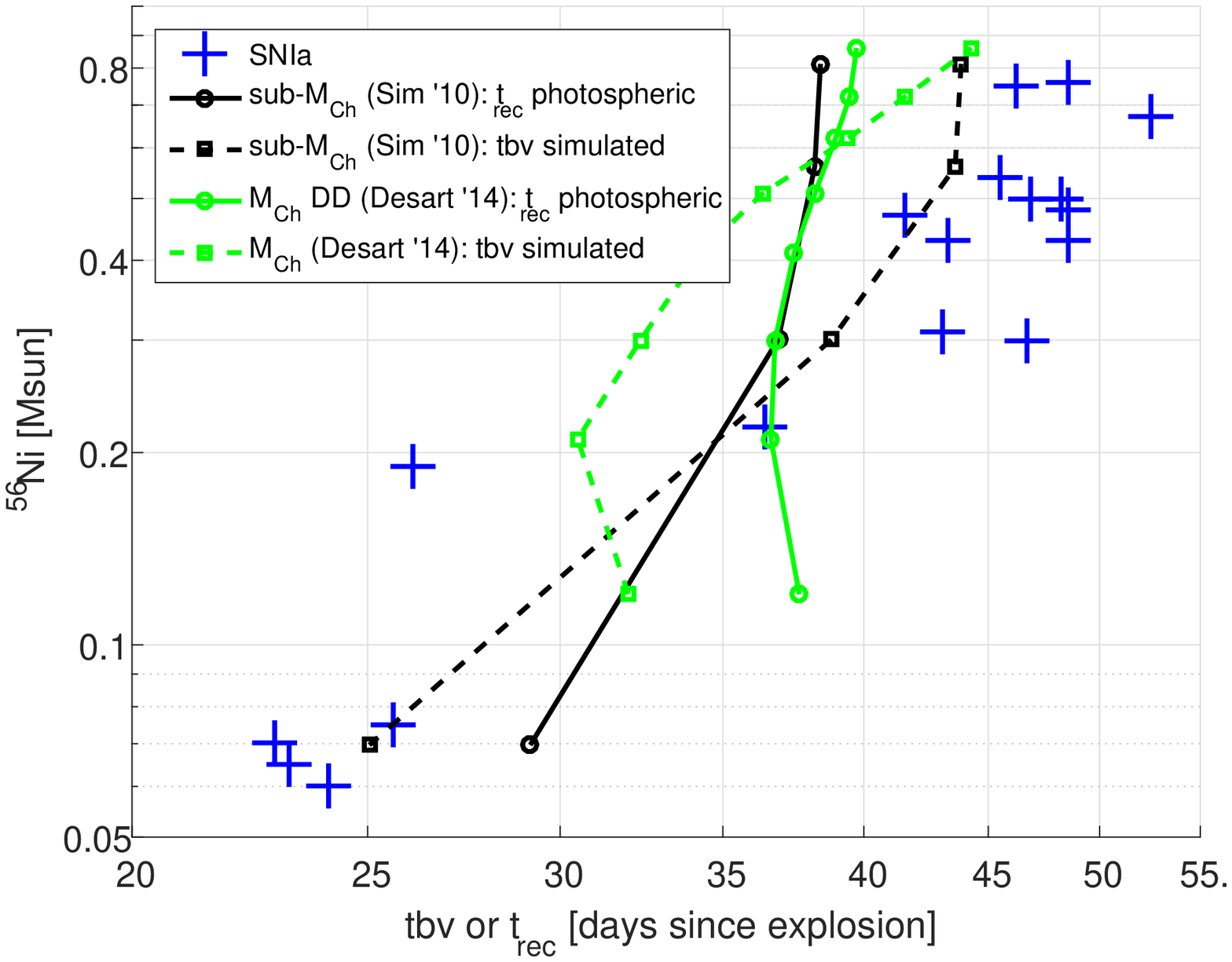}
	\hspace{2mm}
	\includegraphics[width=\columnwidth]{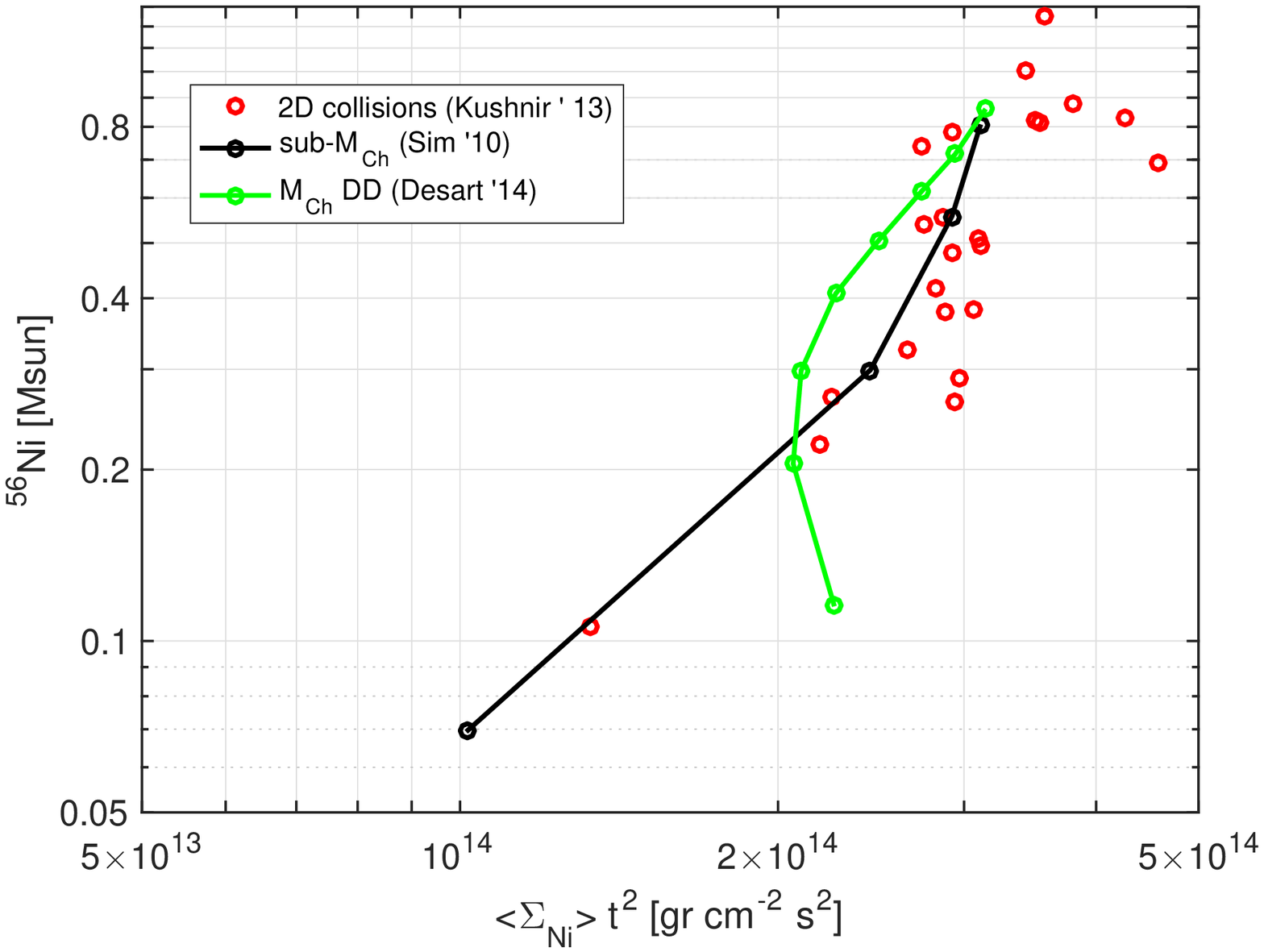}	
	\caption{\textbf{left:} The physical width luminosity relation as \nickel mass versus tbv (inferred from light curves) or recombination time (from photospheric emission model or simulations) for a sample of SNIa (blue) and for ejecta from models of central detonations of sub-$M_{\rm ch}$ WDs (black) or delayed detonation of $M_{\rm ch}$ WDs (green), showing either the photospheric emission results (solid lines) or tbv from the simulated light curve (dashed lines). \textbf{right:} \nickel mass versus \nickel column density for models of central detonations of sub-$M_{\rm ch}$ WDs (black), delayed detonation of $M_{\rm ch}$ WDs (green), and collision scenarios \protect\citep{kk13}}
	\label{fig:trec_models}
\end{figure*}
\begin{figure*}
	\centering
	\includegraphics[width=\columnwidth]{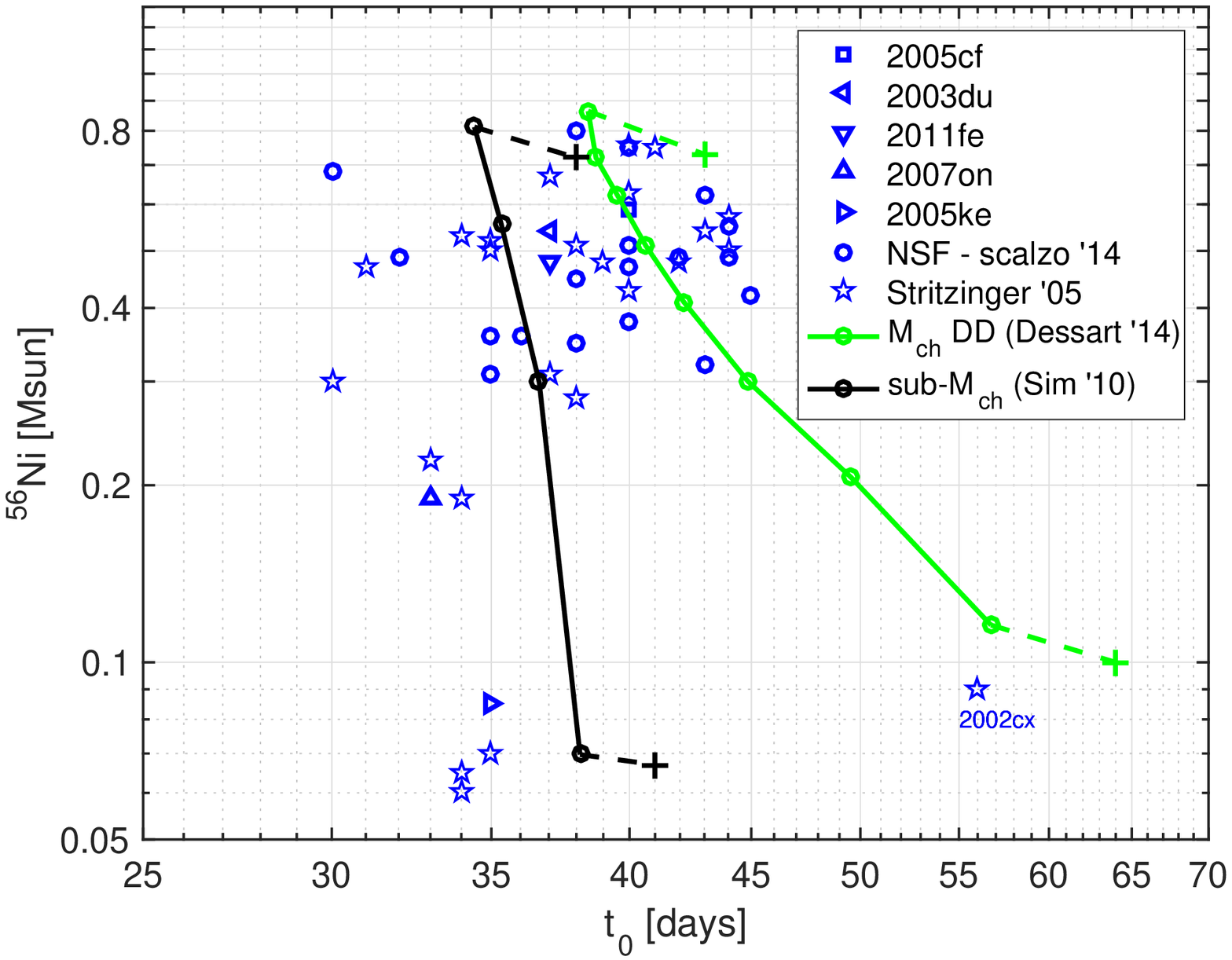}
	\hspace{2mm}
	\includegraphics[width=\columnwidth]{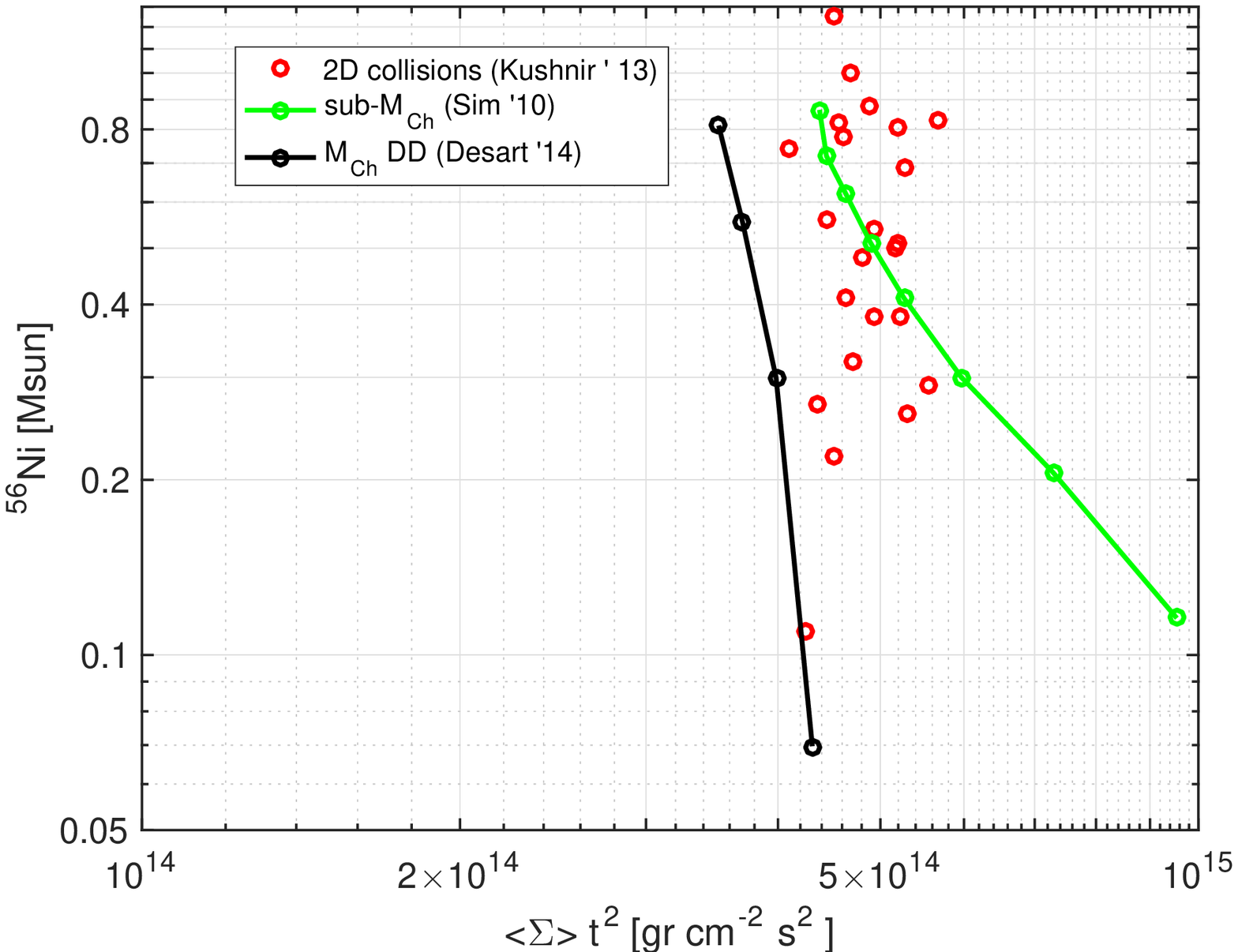}	
	\caption{\textbf{left:} The physical \textbf{bolometric} width luminosity relation as \nickel mass versus $t_0$ for a sample of SNIa (blue) and for ejecta from models of central detonations of sub-$M_{\rm ch}$ WDs (black) or delayed detonation of $M_{\rm ch}$ WDs (green). \textbf{right:} \nickel mass versus total column density (for $\gamma$-rays) for models of central detonations of sub-$M_{\rm ch}$ WDs (black), delayed detonation of $M_{\rm ch}$ WDs (green), and collision scenarios \protect\citep{kk13}}
	\label{fig:t0_models}
\end{figure*}

\subsection{Two physical width-luminosity relations}
The (color) physical WLR, with the $^{56}$Ni mass as the luminosity parameter and tbv (approximately the recombination time) as the width parameter, is shown in the left panel of figure  \ref{fig:trec_models} for several observed supernovae (see appendix~\ref{sec:appendixC} for details of the sample shown). 
The results of radiation transfer calculations (done in this work) for the sub-$M_{\rm ch}$ ejecta from \cite{s+10} and for the $M_{\rm ch}$ delayed detonation ejecta from \cite{dbhk14} are shown for comparison (dashed lines).  For each ejecta, the resulting recombination time based on equation Eq. \eqref{eq:trecequation} (with $f=0.7$) is shown (solid lines). As can be seen both explosion models produce a WLR which is roughly consistent with the observations (the $M_{\rm ch}$ less so). 
The origin of the WLR can be traced to the relation between the $^{56}$Ni column density $\langle \Sigma_{\rm Ni}\rangle$ and mass as shown in the right panel of the figure (using figure \ref{fig:trec_vs_sigmanini} for relating column density to recombination time). \emph{Brighter type Ia's have wider light curves on this relation because higher $^{56}$Ni mass corresponds to higher $^{56}$Ni column densities}. As can be seen in the right figure, the resulting ejecta from 2D calculations of direct collisions of WDs have a $\langle \Sigma_{\rm Ni}\rangle$-$^{56}$Ni relation which is similar to the sub-Chandrasekhar models (see table~\ref{table:ejecta_table_literature}), and are thus similarly consistent with observations. 2D radiation transfer calculations which are beyond the scope of this paper are required to confirm this (see also \cite{rkgr09}).   

A second, bolometric, physical WLR was presented in paper I with the $^{56}$Ni as the luminosity parameter and the gamma-ray escape time $t_0$ as the width (see also \citealt{strit05}). To allow comparison between the two WLRs, we reproduce it here in figure  \ref{fig:t0_models} and present it using similar axis choices as in figure \ref{fig:trec_models}. Observations suggest that  $t_0$ does not change significantly across the brightness range of type Ia. While this is a trivial WLR, it still contains valuable information since not all models reproduce it. Indeed, the $M_{\rm ch}$ delayed detonation models have an inverse WLR where faint explosions have longer gamma-ray escape times and are inconsistent with observations. The bolometric WLR is reflected in the relation between total column density $\langle\Sigma\rangle$ and $^{56}$Ni as can be seen in the right panel. Here the relation is even more direct than that of the color WLR since $t_0$ is a simple function of $\langle\Sigma\rangle$ (see Eq. \eqref{eq:t0Sigma}). \emph{Sub-$M_{\rm ch}$ central ignitions and direct collisions have a trivial bolometric WLR because explosions with varying  $^{56}$Ni masses have the same total column densities. $M_{\rm ch}$ models have longer escape times for dimmer explosions because they have larger total column densities for lower $^{56}$Ni masses.}

\section{Summary and Discussion}\label{sec:summary}

In this work we set to identify quantitative and robust implications that the observed width-luminosity relation has on the ejecta of type Ia supernovae. The luminosity scale has been long known to be proportional to the amount of $^{56}$Ni. The main challenge is the quantitative association of the width to a physical time scale. We have identified two timescales which can be robustly calculated and inferred from observations. The first is the gamma-ray escape time $t_0$, which governs the shape of the bolometric light curve (paper I, \citealt{j99,strit05}). The second is the time of recombination of the $^{56}$Ni decay products from doubly to singly ionized, which governs the color evolution \citep{pi01,kw07,h+17}. 
Using radiation transfer simulations of a wide range of ejecta, we showed that the recombination time is equal to the time tbv where the color B-V curve experiences a sharp break (see figure \ref{fig:tion_vs_tbv}), that has been recently realized  by \cite{burns14} to be an excellent width parameter observationally. 

The physical origin of the break in the B-V light curves has been identified (see figure \ref{fig:no_recombination_effect}). While the singly ionized ions have much higher opacities than the doubly ionized ones, the recombination time reflects a transition from optically thick to optically thin emission. At early stages of recombination, the increase in abundance of singly ionized ions compensates for the decrease in opacity due to the declining temperatures. Once a significant fraction of the ions have recombined (the recombination time) there is no more new sources of high opacity and a sharp drop is encountered. 

A simple quantitative model for estimating the recombination time was derived in \S\ref{sec:tbv_predicted} and it was shown that the recombination time mainly depends on the $^{56}$Ni column density $\langle \Sigma_{\rm Ni}\rangle$ (see Eqs. \eqref{eq:trecequation} and \eqref{eq:SigmaNiNi},  and figure \ref{fig:trec_vs_sigmanini}).

The two time-scales are thus set by two column densities of the ejecta - the recombination time by the $^{56}$Ni column density and the gamma-ray escape time, $t_0$, by the total column density. Two physical width-luminosity relations are thus obtained that reflect the relation between the $^{56}$Ni mass and the two column densities (see figures \ref{fig:trec_models} and \ref{fig:t0_models}). 
By comparing representative ejecta that result from hydrodynamic simulations it is shown that the sub-$M_{\rm ch}$ explosions resulting from central ignition of WDs by \cite{s+10} and the direct collision simulations by \cite{kk13} have similar column densities that are consistent with observations. Chandrasekhar mass, delayed detonation explosion calculations from \cite{dbhk14} fail to reproduce the bolometric WLR at the faint end (paper I, \citealt{kk13,s+14}). As shown in paper I, this is a generic feature of Chandrasekhar mass models and is unlikely to be resolved by modifications to the explosion mechanism.

While it has been demonstrated here that there are two physical times scales that affect the light curves, it is natural to ask if the two timescales are sufficient to set the temporal evolution. In other words, could there be more time-scales involved?

\begin{figure*}
	\includegraphics[scale=0.48]{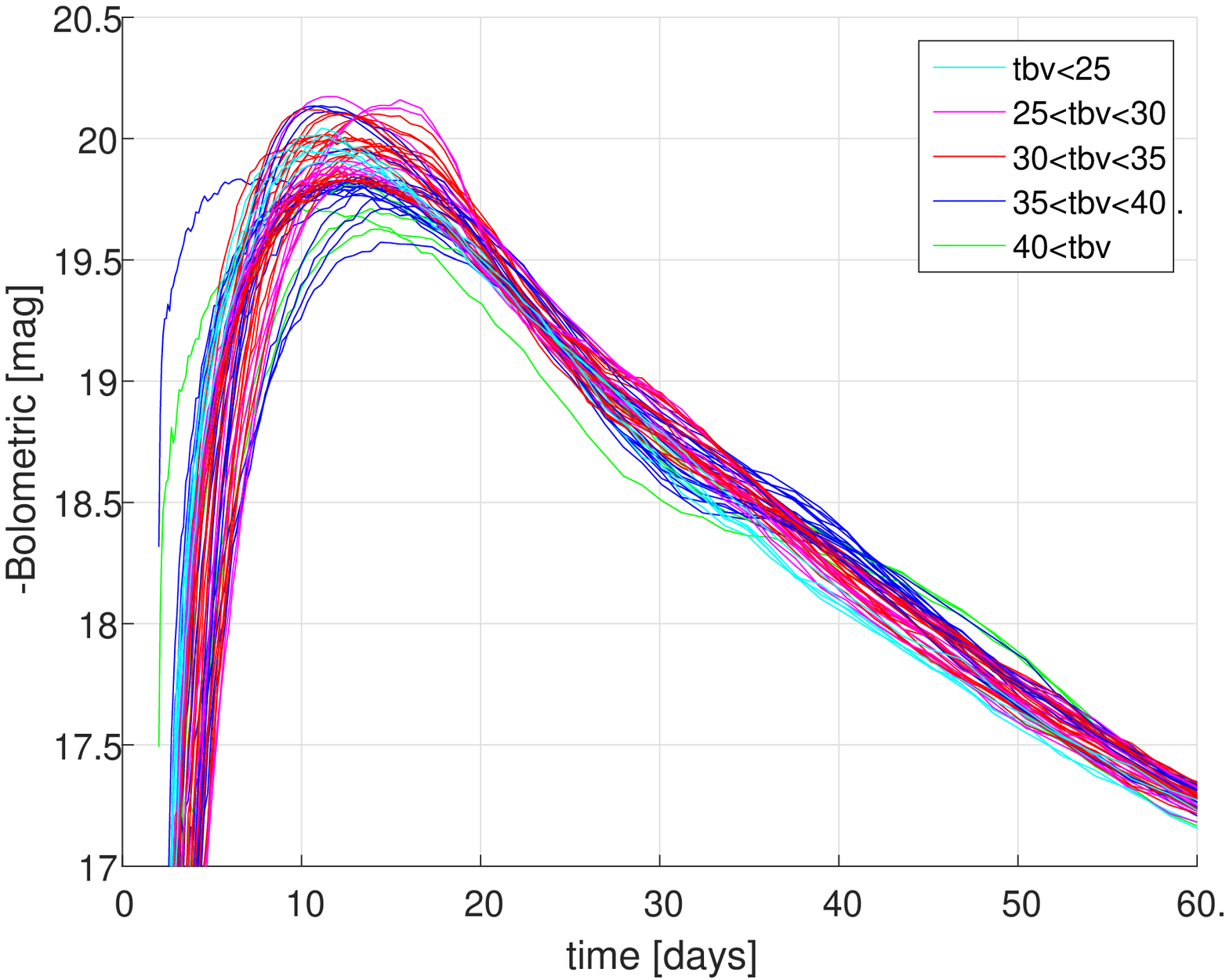}
	\hspace{2mm}
	\includegraphics[scale=0.48]{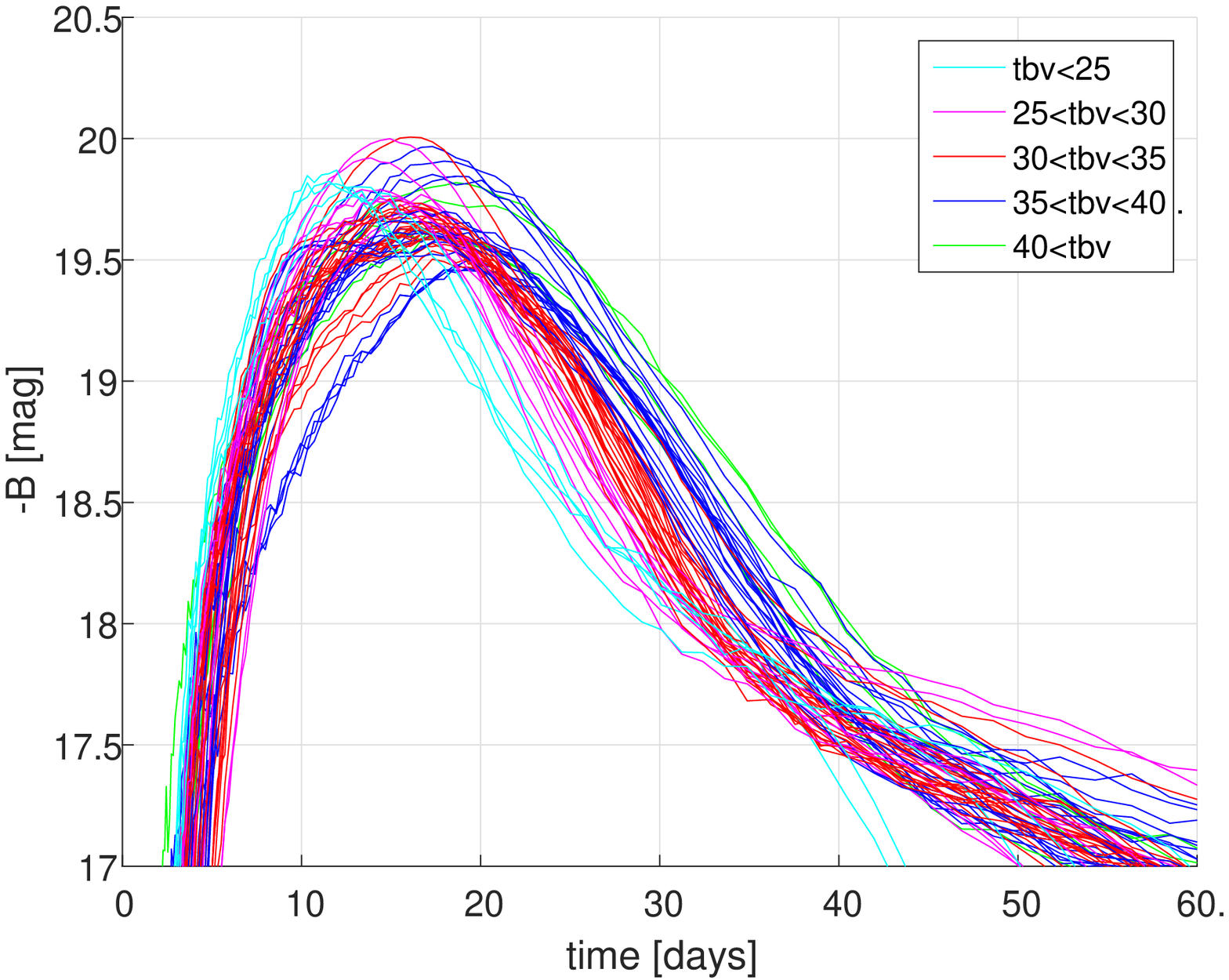}
	\vspace{2mm}
	\includegraphics[scale=0.48]{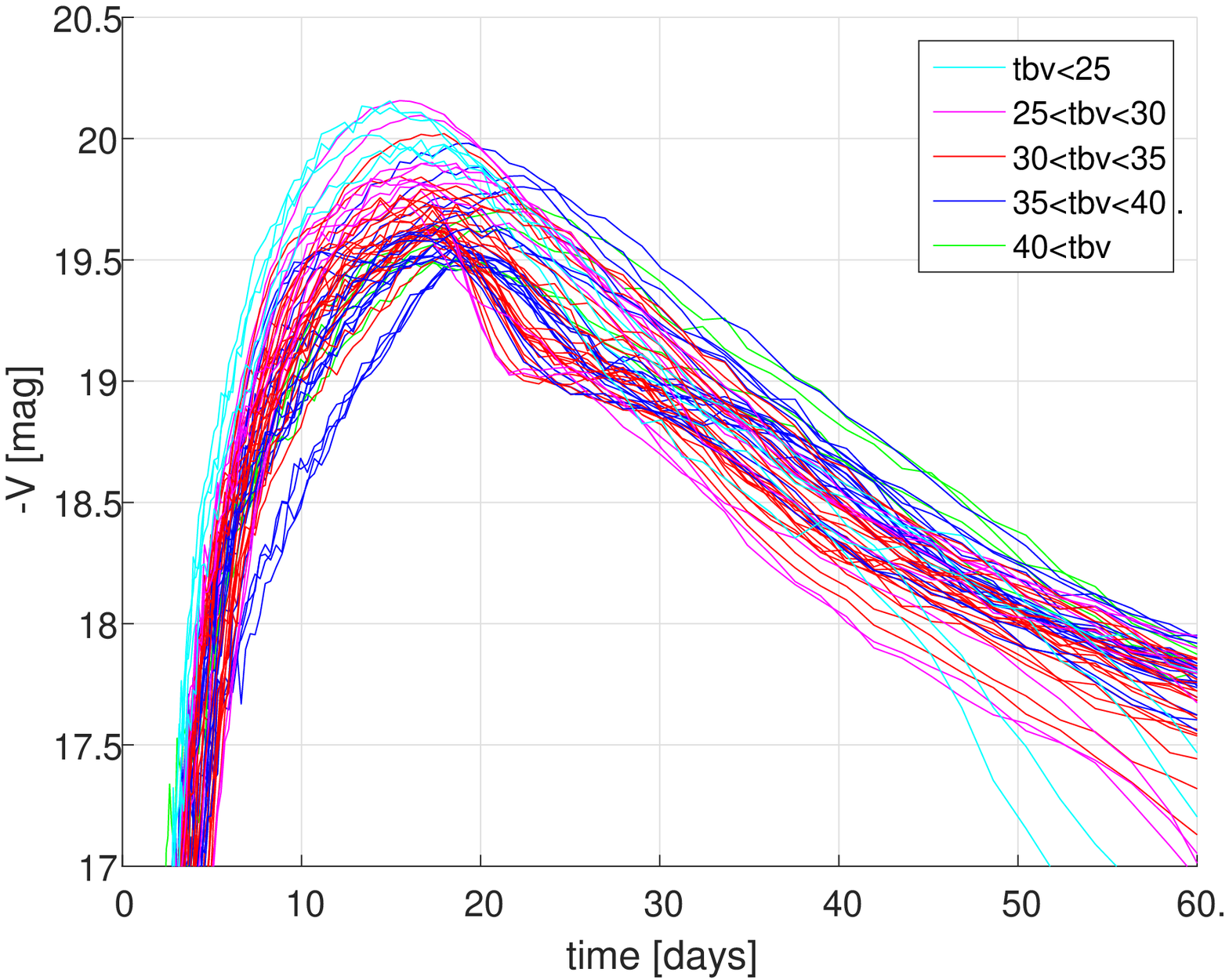}
	\hspace{2mm}
	\includegraphics[scale=0.48]{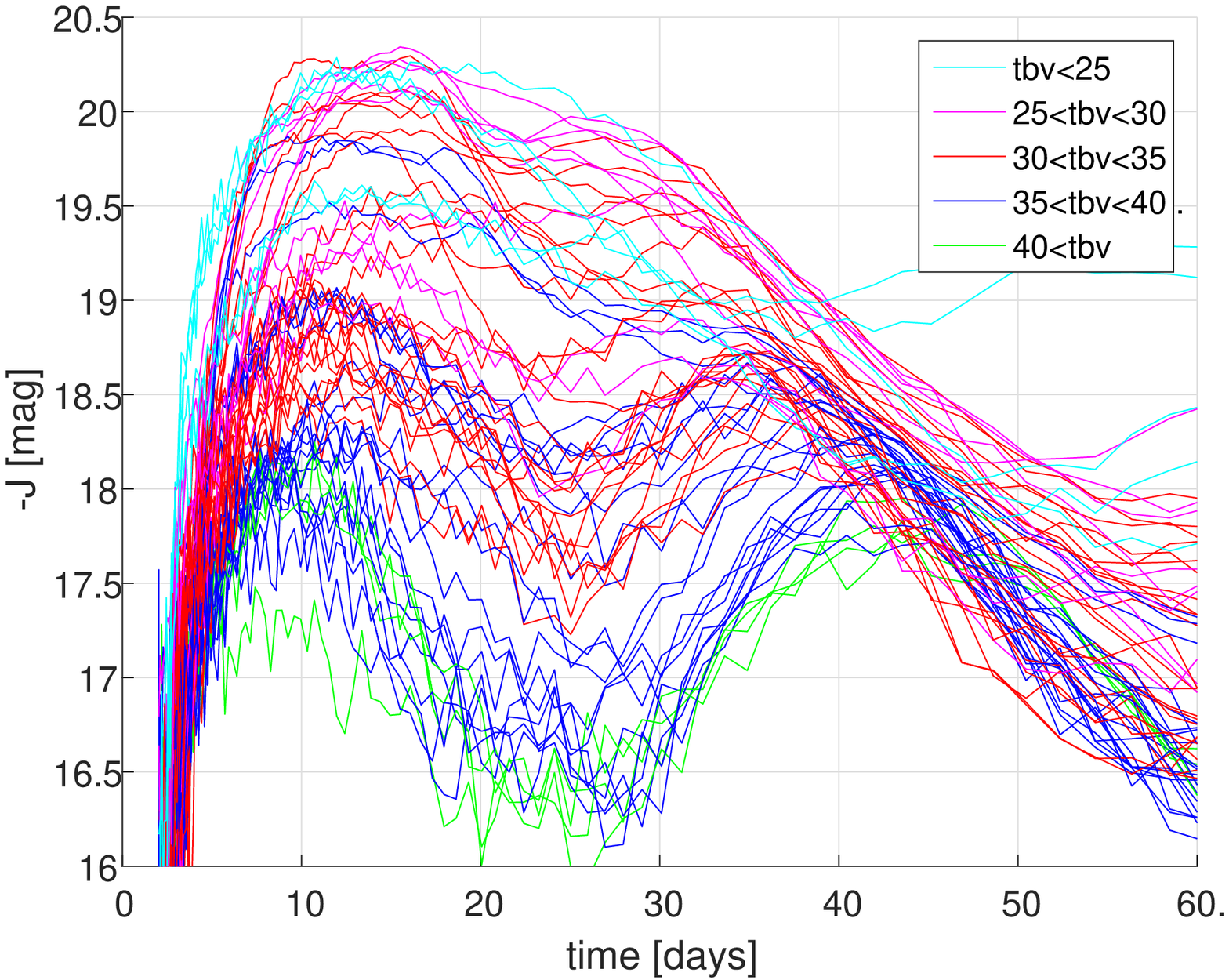}
	\caption{light curves simulated for a set of synthetic ejecta, all with $t_0=35$ days, normalized to a \nickel mass of $0.7M_{\odot}$, and colored by order of their tbv: $\rm tbv<25$ (cyan), $25<\rm tbv<30$ (magenta), $30<\rm tbv<35$ (red), $35<\rm tbv<40$ (blue), and $40<\rm tbv$ (green). Plotted are the bolometric light curve (top left panel), B-band (top right panel), V band (bottom left panel) and J band (bottom right panel) light curves. }
	\label{fig:sample_by_sbv}
\end{figure*}

This question is partly answered in figure~\ref{fig:sample_by_sbv}, where simulated light curves for various ejecta (all with $t_0=$35 days) are plotted, normalized by the \nickel mass. The light curves are presented in colors that represent the value of their tbv. The bolometric light curves are all tightly clustered at the relevant times. The B-band light curves are well ordered by their tbv values for a large period of their post-maximum decline, as well as the V-band light curve (although less tightly so). This is especially true for the post maximum decline. The IR light curves, however, feature a much wider range of luminosities as shown for example through the J-band in figure~\ref{fig:sample_by_sbv}.  
It thus seems that the optical post maximum decline is mostly governed by $t_0$ and tbv, while the early rise, as well as the IR light-curves, seem to involve additional time scales and are thus likely to contain more information about the explosion mechanism.

\section*{acknowledgments}
We thank Mark Phillips for the suggestion to focus our studies on the B-V break as a width parameter for the WLR which is at the basis of this work. We thank Doron Kushnir, Subo Dong, Eli Waxman, Avishay Gal-Yam, Ehud Nakar and Stuart Sim for useful discussions. We thank Masaomi Tanaka for providing the collection of benchmark calculations. This work was partially supported by the ICORE Program (1829/12) and the Beracha Foundation.

\appendix
\section{Radiative Transfer Code, description and validation}\label{sec:appendixA}

In this work, we have used "URILIGHT", a new Monte-Carlo radiative transfer code written in FORTRAN 90 which is now public and can be obtained here: \\
https://www.dropbox.com/sh/kyg1z1xwi0298ru/ AAAqzUMbr6AkoVfkSVIYChTLa?dl=0 \\
The code is described in \S\ref{sec:CodeDescription}, convergence tests are presented in \S\ref{sec:CodeConvergence} and validation by comparison to previous calculations is presented in \S\ref{sec:CodeValidation}.
\subsection{Description of the code}\label{sec:CodeDescription}
The physical assumptions underlying the code are mostly similar to the SEDONA program following \citep{ktn06}, as described in section \S~\ref{sec:radiative_transfer} and detailed below. 

\subsubsection{$^{56}$Ni decay chain and gamma-ray transfer}
The 1D homologous ejecta is read from external files or created for simple configurations and converted to arrays that contain  the mass in each cell, velocity of the edges and the isotope composition. The decay chain of $^{56}$Ni$\rightarrow^{56}$Co$\rightarrow ^{56}$Fe and the gamma-ray transfer are calculated first. Half lives of $6.075\rm d$ and $77.233 \rm d$ are assumed for $^{56}$Ni and $^{56}$ Co respectively \cite{milne04}, with gamma-ray energies and branching ratios taken from \cite{ambwani88}, including the pair annihilation line at $0.511\rm MeV$. Positrons are assumed to be produced with a branching ratio of $0.19$ of $^{56}$Co decays with an average kinetic energy of $0.632\rm MeV$ which is assumed to be deposited locally immediately followed by immediate annihilation of the positrons. Gamma-ray packets are randomly created in each cell in proportion to the $^{56}$Ni mass and the branching ratios. Two types of interaction with matter are included - Compton scattering and photoelectric absorption. Compton scattering is calculated by treating all electrons (bound and unbound) as free and cold, using the Klein -Nishina cross section with incoming and outgoing photon energies and directions calculated exactly. Photoelectric  absorption is restricted to the K-shell electrons using the approximate cross section per ion \citep{ktn06},
\begin{equation}
\alpha=\sigma_T\alpha_e^48\sqrt{2}x^{-3.5}Z^5,
\end{equation}
where $\sigma_T$ is the Thompson cross section, $\alpha_e$ the fine structure constant, $Z$ the charge number of the element and $x=h\nu/m_ec^2$ is the energy of the gamma-ray photon in units of the rest mass energy of the electron. The energy deposited by the gamma-rays (in Compton scatterings and photo-absorption) and positrons are separately recorded in each cell for each of the time steps that are later used in the UVOIR calculation.
\subsubsection{UVOIR transfer}
The calculation of the transfer of UVOIR is preformed after the decay+gamma-ray calculation has been completed.  UVOIR radiation is treated using energy packets that all have the same initial energy that are created as a function of time and position in the ejecta in numbers that are proportional to the energy deposition from the gamma-rays and positrons. The wavelength of each packet is chosen randomly with a distribution that is proportional to the local emissivity.  The wavelength of packets is constant in the lab frame but changes during interactions. The ionization state and energy-level occupation of the ions at each point in the ejecta is assumed to be in local thermal equilibrium (LTE) and set by the plasma temperature at that point. The temperature profile and UVOIR transfer are calculated in discrete time steps that each involve iterations of transfer and temperature calculations. In the beginning of each time step, the temperature profile from the previous time step is assumed as a first guess. The transfer of UVOIR energy packets that already exist, as well as new UVOIR packets based on the gamma-ray+positron deposition are calculated throughout the time-step using the opacities set by the temperature (described below). During the propagation of each photon through a spatial cell, its energy deposition in the plasma in the cell is calculated based on the amount of length traversed  by the photon rather then by its interactions \citep{lucy05a}. Once the transfer calculation is done, the temperature of each cell is solved-for by requiring thermal balance: that the (temperature dependent) emissivity equals the sum of the UVOIR absorption by the photons that traversed the cell and the gamma-ray and positron deposition. A second iteration begins where the transfer of UVOIR is recalculated using the new opacities followed by a second calculation of the temperature. These two iterations are sufficient as shown below in \S\ref{sec:CodeValidation}.

The opacities are calculated on a grid of wavelengths with equal spacing $d\lambda$ ($d\lambda=10$\AA) and are separated into scattering (frequency in the plasma frame does not change) and absorption (frequency in the plasma frame is chosen with probability distribution proportional to the emissivity). In all types of interaction the new orientation is chosen randomly, isotropically in the plasma frame. The wavelength dependent thermal emissivity is calculated using the absorptive opacity. Bound-bound transitions are the main source of scattering and absorption (calculation described below). Thompson scattering on the free electrons and free-free absorption are included. Bound-free interactions are ignored.

Free-free absorption is calculated approximately with the absorption coefficient per ion given by \citep{rybickylightman86}
\begin{equation}
\alpha_{\rm ff}=\frac{4e^6}{3m_ehc}\left(\frac{2\pi}{3m_e kT}\right)^{1/2}n_e\sum_i z_i^2n_i\frac{1-e^{-h\nu/kT}}{\nu^3}  
\end{equation}  
where $n_e$ is the density of free electrons and $z_i$ and $n_i$ are the ionization level (charge) and density of ion $i$ respectively. Note that the gaunt factor is taken to be unity.

Bound bound transitions are treated with the expansion opacity formalism \citep{Karp77,EastmanPinto93}. As a photon propagates through the expanding ejecta, its plasma-frame wavelength $\lambda$ is redshifted continuously  When the wavelength of a photon crosses the wavelength of a line $i$,  the probability that the photon interacts with a line  is $1-e^{-\tau_i}$, where $\tau_i$ is the optical Sobolev optical depth, 
\begin{equation}\label{eq:tausobolev}
\tau=\frac{\pi e^2}{m_ec}fn_l\lambda t (1-e^{-h\nu/ kT})
\end{equation}
where $f$ is the oscillator strength, $n_l$  is the density of the ions in the lower state, $t$ is the time since explosion and $\nu=c/\lambda$.  The opacities are calculated on a grid of wavelengths (with equal spacing $d\lambda=10$\AA) and the interaction coefficient in each wavelength bin is calculated by summing the contribution of the lines in the bin, 
\begin{equation}\label{eq:alphabbtot}
\alpha_{\rm bb,tot}=\frac{1}{d\lambda~ ct}\sum_{i} \lambda_i (1-e^{-\tau_i}).
\end{equation}
Following an interaction it is assumed that there is a chance $\epsilon$ (which is a global parameter of the simulation \citealt{nugent97,ktn06}) that the photon is re-emitted at a different transition and such an event is treated as absorption. There is a probability $1-\epsilon$ that the photon is re-emitted at the same wavelength (within the line width). In that case, there is a probability 
\begin{equation}
\beta=\frac{1-e^{-\tau}}{\tau}
\end{equation}
that the photon escapes and such an event is treated as scattering. There is however a probability of $1-\beta$ that the photon has a new interaction, which can then end up as absorption,  scattering, or yet another interaction. Such multiple interactions are assumed to occur instantaneously. The overall probability for absorption is the sum of an infinite geometric series with ratio $(1-\epsilon)(1-\beta)$ and is given by:
\begin{equation}\label{eq:pabs}
p_{\rm abs}=\frac{\epsilon}{1-(1-\epsilon)(1-\beta)}
\end{equation}
while the probability for an eventual scattering is $1-p_{\rm abs}$. The total bound-bound opacity, Eq. \eqref{eq:alphabbtot} is thus separated to absorption and scattering \citep[equation 8 in][]{ktn06}, 
\begin{align}\label{eq:alphabbabssca}
\alpha_{\rm bb,abs}&=\frac{1}{d\lambda~ ct}\sum_{i} \lambda_i p_{\rm abs,i}(1-e^{-\tau_i}),\cr 
\alpha_{\rm bb,sca}&=\frac{1}{d\lambda~ ct}\sum_{i} \lambda_i (1-p_{\rm abs,i})(1-e^{-\tau_i}),\cr 
\end{align}
where $\tau_i$ and $p_{\rm abs,i}$ are calculated for each line using equations \eqref{eq:tausobolev}, \eqref{eq:pabs}.

It was demonstrated in \citep{ktn06} that this formalism results in small errors in the light curves and in the spectrum compared to individual line modeling, as long as the wavelength binning is small enough ($\sim10$\AA) and $\epsilon$ is set to relatively high values ($0.3-1$). Unless otherwise specified, the absorption probability per interaction is set to $\epsilon=0.8$ following \citep{kw07}.

\subsection{Convergence}\label{sec:CodeConvergence}

\begin{figure*}
	\centering
	\includegraphics[width=\columnwidth]{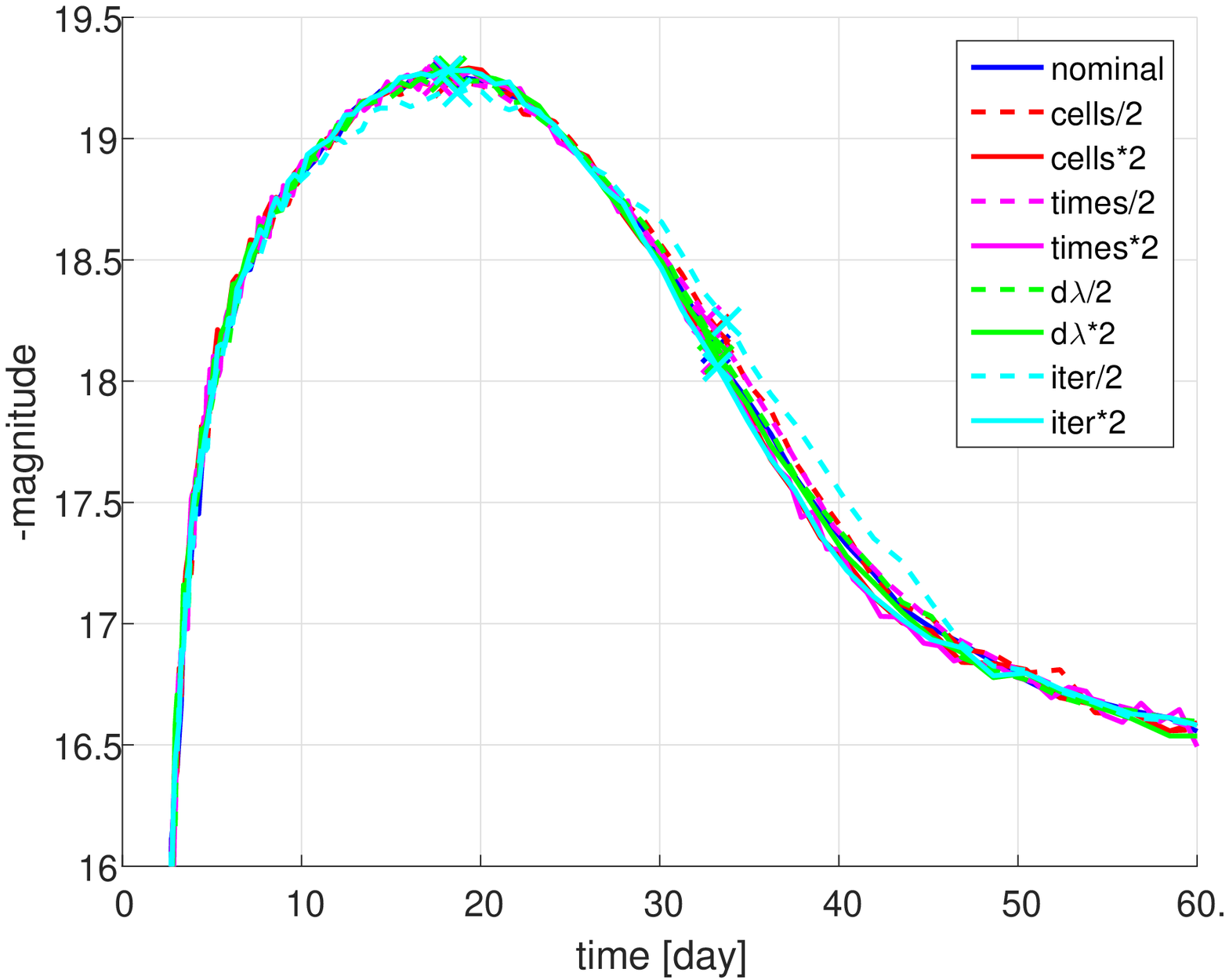}
	\hspace{2mm}
	\includegraphics[width=\columnwidth]{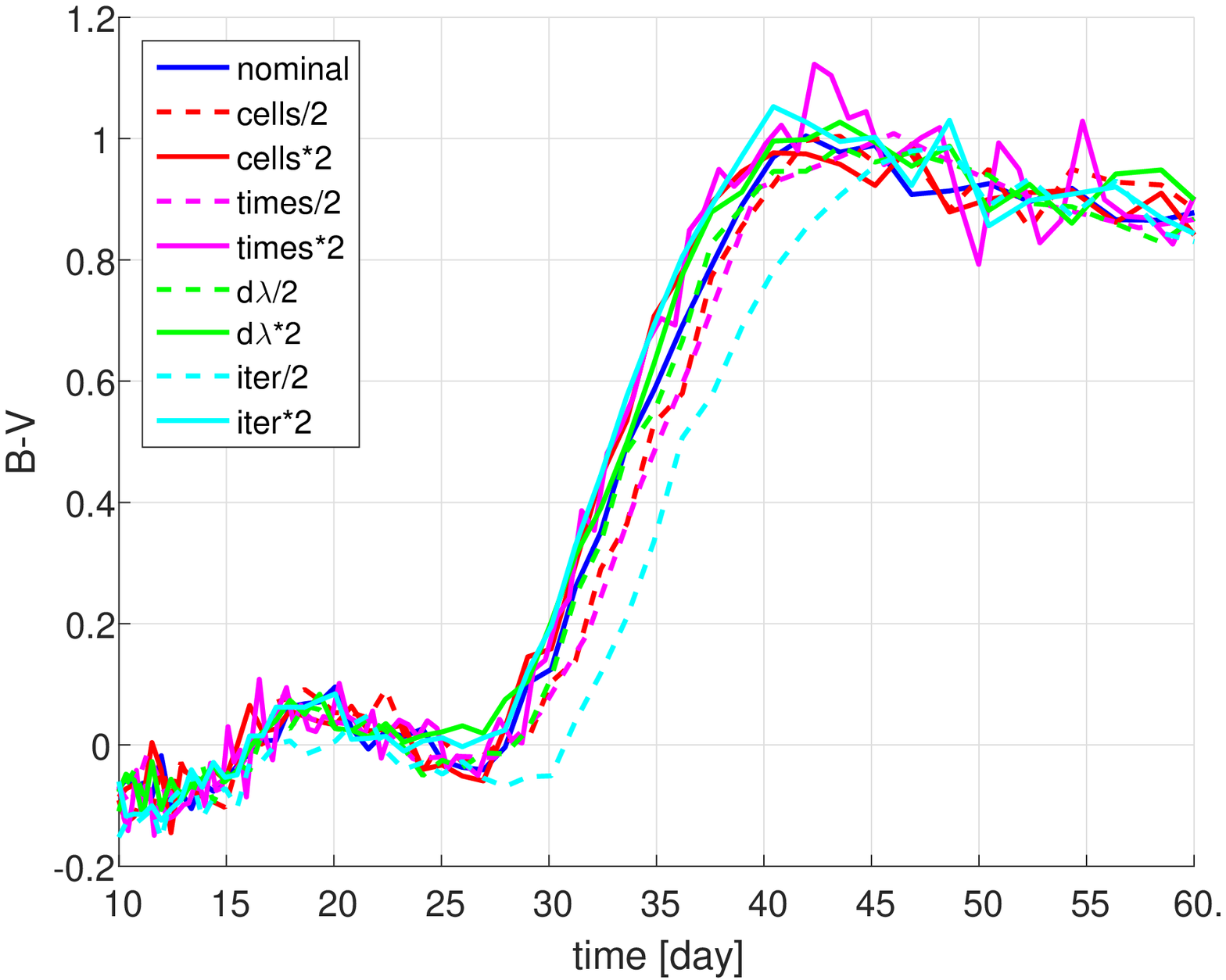}	
	\caption{Numerical convergence tests for B band (left figure) and B-V (right figure), for a synthetic ejecta with $M(^{56}\rm Ni)=0.7M_{\odot}, t_0=35$ days: nominal run (blue) vs. factor 0.5 (dashed red) and 2 (solid red) in spacial resolution, factor 0.5 (dashed magenta) and 2 (solid magenta) in time resolution, factor 0.5 (dashed green) and 2 (solid green) in spectral resolution, and half (dashed cyan) and twice (solid cyan) the number of iterations on the temperature profile.}
	\label{fig:numerics3}
\end{figure*}

The simulations presented in the current work were performed with the following numerical specifications: 120 spacial cells (except for simulations of ejecta from other works, which were performed with the number of cells provided in these works), 100 temporal steps logarithmically distributed between 2 and 80 days, and a spectral grid with constant wavelength spacing of $10$\AA. In addition, the number of iterations for setting the temperature profile was set to 2 (rather than a varying convergence criterion). Figure~\ref{fig:numerics3} shows, for a given ejecta ($0.1M_{\odot}$ stable iron, $0.7M_{\odot}$ of \nickel, $t_0=35$ days, $E_K=1.5\cdot10^{51}$ ergs, and $0.3M_{\odot}$ of IMEs with no mixing) that the simulations are converged in all these parameters.

\subsection{Comparisons with previous results} \label{sec:CodeValidation}
A comprehensive comparison of several radiative transfer codes for SNIa models was done in \cite{tanaka13}, and we follow along this work for comparing our results using the radiative transfer program URILIGHT to previous radiative transfer codes for two scenarios.

\begin{figure*}
	\centering
	\includegraphics[width=\columnwidth]{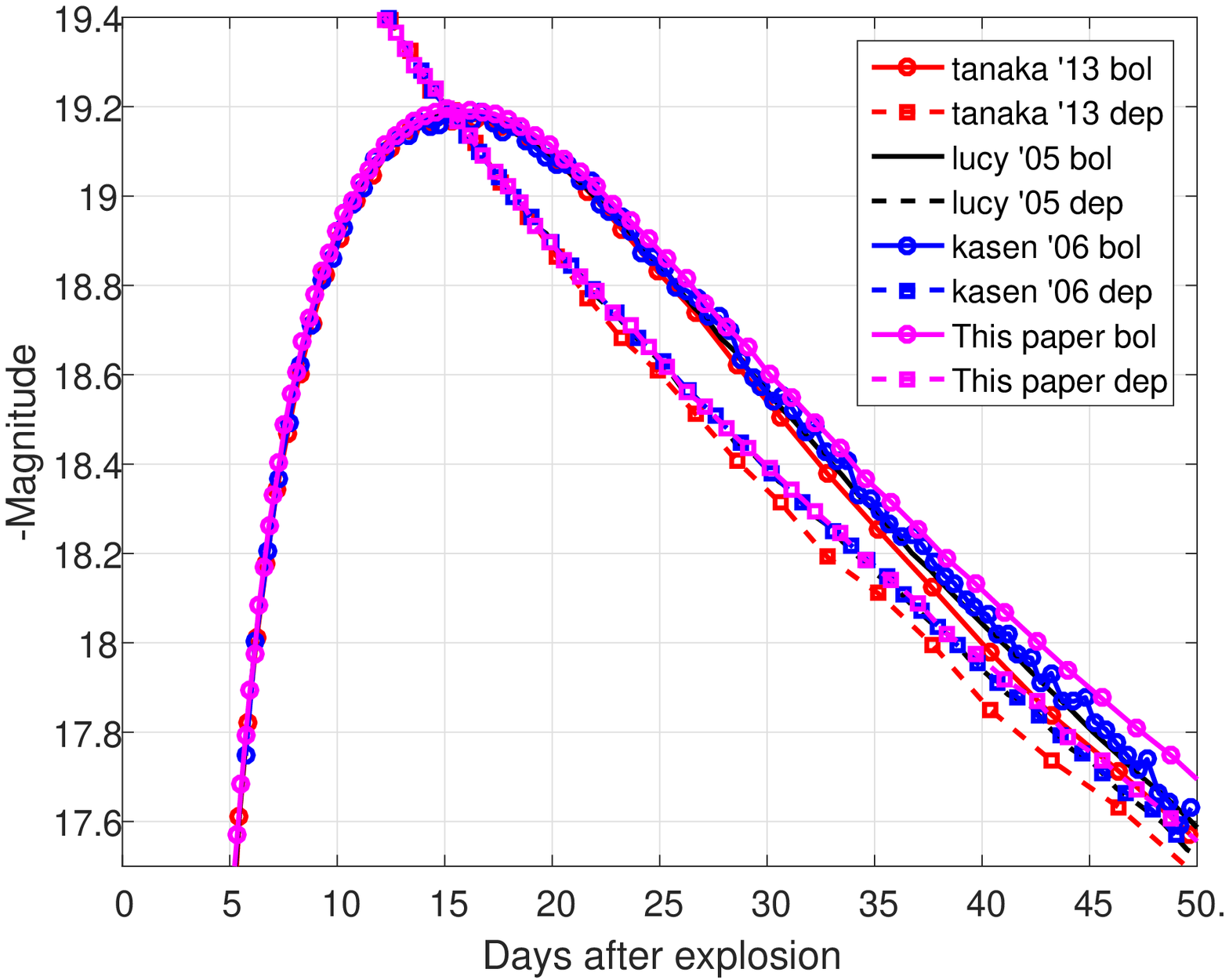}
	\hspace{2mm}
	\includegraphics[width=\columnwidth]{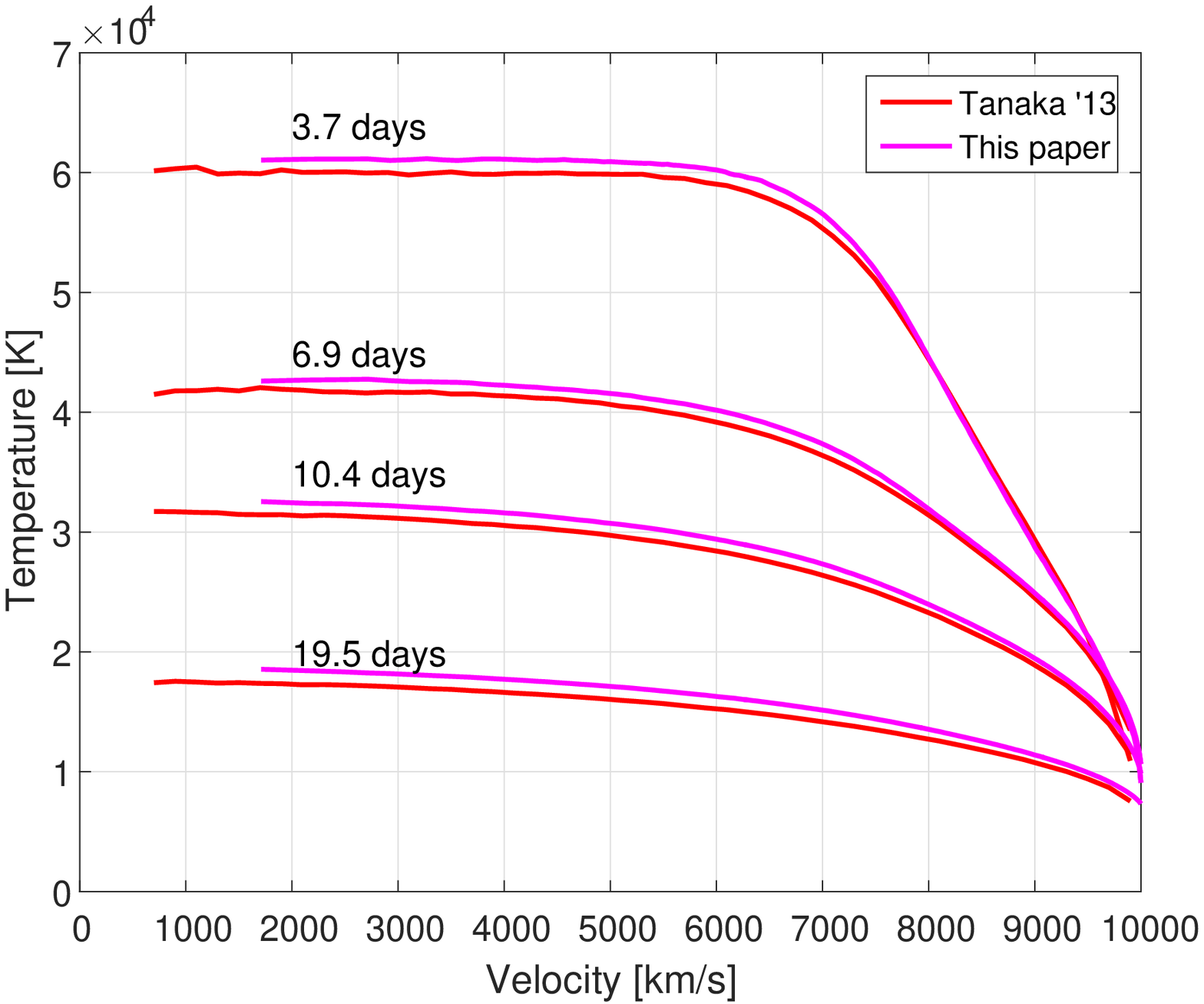}	
	\caption{Comparison of several radiative transfer solutions to the simplified SNIa model of \protect\cite{lucy05a}. \textbf{left:} Bolometric light curves (solid lines) and deposition rate (dashed lines) for the moment equation solution of \protect\cite{lucy05a} (black), and the Monte-Carlo solutions of \protect\cite{tanaka13} (red), \protect\cite{ktn06} (blue) and the current work (magenta). \textbf{right:} Temperature profiles of the ejecta at different times from \protect\cite{tanaka13} (red) and from this work (magenta). }
	\label{fig:lucy_comparison}
\end{figure*}

The first configuration is the simple SNIa model proposed in \cite{lucy05a}, consisting of a uniform density ejecta with mass $1.39 M_{\odot}$, maximal velocity of $10^4$ km/s, and pure \nickel from the center up to $0.5 M_{\odot}$, which then drops linearly (in mass coordinate) to zero at $0.75 M_{\odot}$ for a total \nickel mass of $0.625 M_{\odot}$. The opacities for this problem are taken as gray opacities with $\kappa=0.1 \rm cm^2/gr$. Figure~\ref{fig:lucy_comparison} compares the deposition rate and the bolometric lightcurves as calculated in \cite{lucy05a}, \cite{tanaka13} and \cite{ktn06} with the current work (the latter two being MC codes with similar logic to URILIGHT in the current work), as well as comparing snapshots of the temperature profiles between \cite{tanaka13} and the current work. The overall results are very close in the various simulations (even more so between URILIGHT and SEDONA), with slightly higher bolometric luminosities in URILIGHT at late times. 

\begin{figure*}
	\centering
	\includegraphics[width=\textwidth]{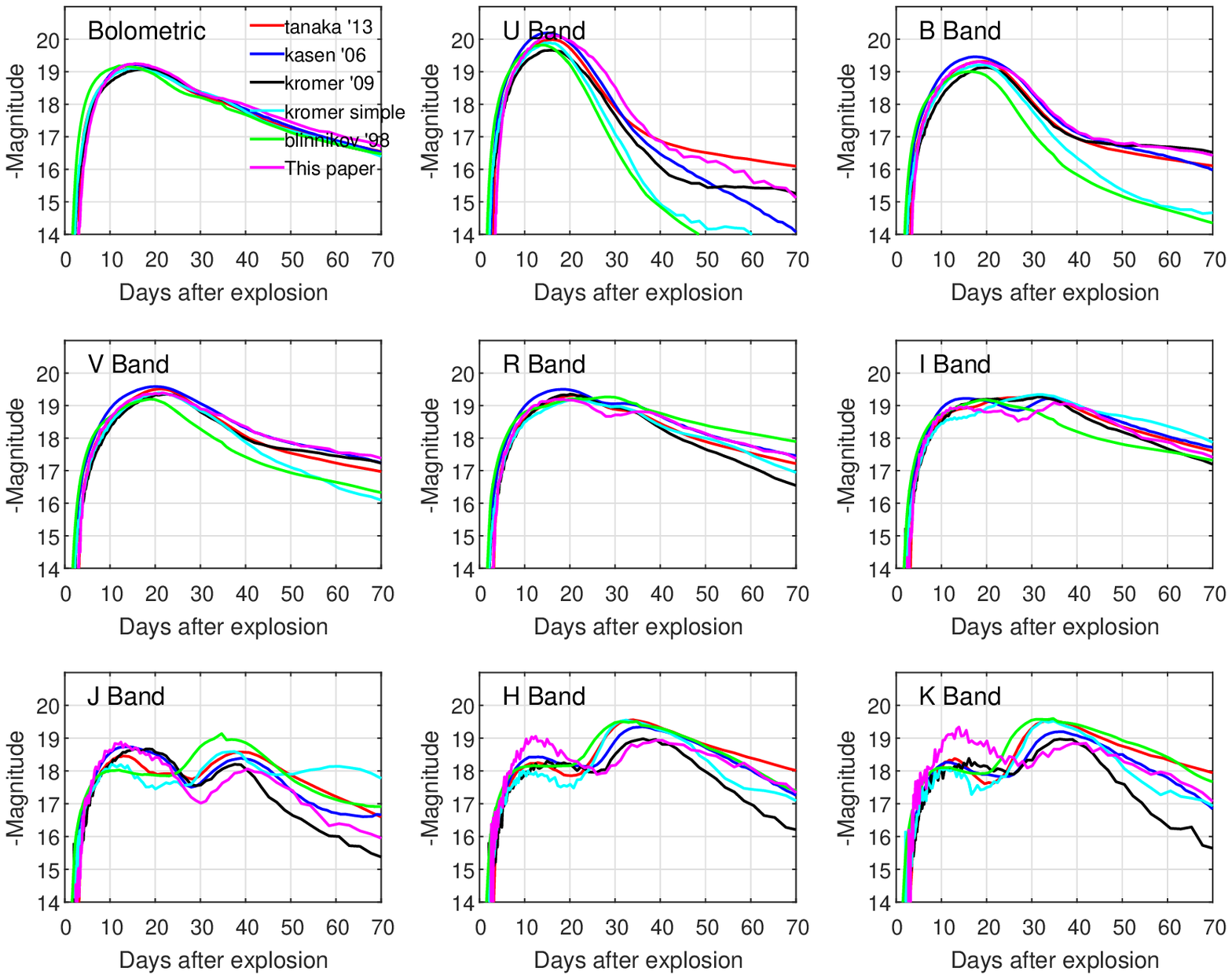}
	\caption{Comparing bolometric and UBVRIJHK light curves for the W7 model \protect\citep{nty84} for several radiative transfer codes: \protect\cite{tanaka13} (red), SEDONA by \protect\cite{ktn06} (blue), ARTIS by \protect\cite{ks09} (black, and their simplified model - cyan), STELLA by \protect\cite{b+98} (green) and URILIGHT of this work (magenta). Note that the simulated lightcurves in this figure are identical to those in figure 13 of \protect\cite{tanaka13}, with the exception of the results of URILIGHT which have been added here.}
	\label{fig:w7_all_bands}
\end{figure*}

\begin{figure}
	\includegraphics[width=\textwidth]{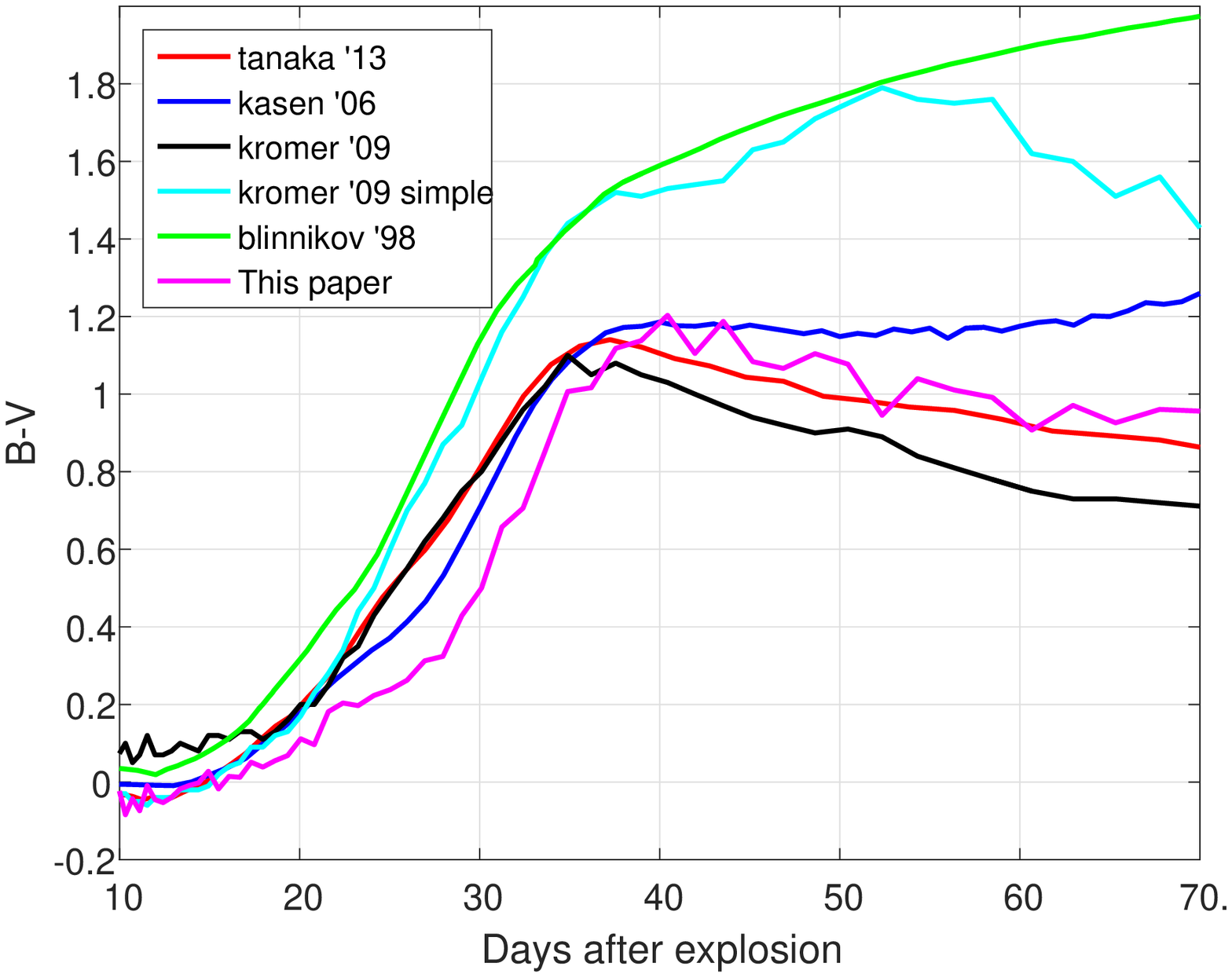}
	\caption{B-V curves for ejecta W7 \protect\cite{nty84} for the same simulations as in figure~\ref{fig:w7_all_bands}.}
	\label{fig:w7_bmv}
\end{figure}

The second configuration is the W7 model for SNIa \citep{nty84}, using multi frequency UVOIR transfer (in our simulations, the full list of \citealt{kurucz94}, CD1, has been used). Figure~\ref{fig:w7_all_bands}, which contains the same numerical results as presented in figure 13 of \cite{tanaka13} in addition to our simulation results, compares bolometric and UBVRIJHK light curves for six different codes: \cite{tanaka13} and SEDONA from \cite{ktn06}, which are both MC codes which assume that level populations as well as the ionization structure is in LTE, but do not require the radiation field to be in equilibrium, ARTIS from \cite{ks09}, also a MC code, both in its full version which calculates excitation states without assuming LTE and in its simplified version (which is fully LTE), STELLA from \cite{b+98} which is a hydrodynamic code with multigroup radiative transfer (also fully LTE), and URILIGHT used in this work. For the bolometric, as well as the optical bands, our results are similar to those of previous works (in the optical range, the results of the fully LTE codes - the simplified version of ARTIS, and STELLA - are systematically different from the others). The color curve B-V is shown in figure~\ref{fig:w7_bmv}, showing that the break time is within 2-3 days of each other for URILIGHT, SEDONA, ARTIS and \cite{tanaka13}. hl{Note that in fully LTE codes, the B-V curve is qualitatively different, and the typical break is not clearly observed.} In the far IR, our results are somewhat different. We note that the particular sensitivity of this region of the spectrum has been previously noted, see for example \cite{k06}.

\section{simulated ejecta configurations}\label{sec:appendixB}
The synthetic ejecta configurations whose results are presented in this work are detailed in this section. By 'synthetic' we mean that these configurations were not drawn from simulations of the explosion mechanism.
In order to link tbv with the physics of the ejecta, we simulated many types of ejecta and looked at the emerging light curves. The ejecta that we simulated are detailed in table~\ref{table:ejecta_table}, and we briefly describe them here: They had, for the most part, $t_0=35$ days, and an exponential density profile as in \cite{wkbs07}. We simulated ejecta with $M_{\rm Ni}=0.7M_{\odot}$ and $M_{\rm Ni}=0.3M_{\odot}$, which had also a central core of $0.1 M_{\odot}$ of Fe, and ejecta with $M_{\rm Ni}=0.1M_{\odot}$ without Fe. The outer layer of these ejecta were always pure carbon. Following the observations that the range of observed $t_0$ is quite narrow, we mostly kept $t_0$ constant for the configurations we simulated (though at 35 days rather than the value of 40 days which would better fit the observations), although we checked the effect of varying $t_0$ from 25 to 45 days. We also checked the effect of varying the ejecta kinetic energy and total mass, of having a constant density profile, of mixing the layers described above (using a moving box average, as described in \cite{wkbs07}), of not having the central iron core (or of adding it, for the $M_{\rm Ni}=0.1M_{\odot}$ ejecta), and of changing the outer layer composition by either adding $0.3 M_{\odot}$ of IMEs or having an equal mixture of oxygen and carbon. We also simulated 'Kasen-like' series, based on the ejecta structures proposed in \cite{kw07} (though not identical to them, since these configurations had $t_0=46$ days which is to high relative to the observed values), i.e. ones in which the basic ejecta has an $0.1 M_{\odot}$ Fe layer, then $0.7 M_{\odot}$ Ni and an outer C layer, and for which lower masses of Ni are achieved by uniformly mixing the Ni layer with IMEs. We also simulated these series with carbon mixed in the Ni instead of IMEs, and without the central core of Fe.

\begin{table*}
	\caption{various synthetic ejecta with simulated light curve, discussed in text and in figures~\ref{fig:tion_vs_tbv},~\ref{fig:Lion_over_Qdep_vs_tion}, ~\ref{fig:Ltbv_vs_photospheric}, and~\ref{fig:trec_vs_sigmanini}.}
	\label{table:ejecta_table}
	\begin{tabular}{|c| >{\centering\arraybackslash}m{11mm} | >{\centering\arraybackslash}m{10mm} | >{\centering\arraybackslash}m{21mm} |c|c| >{\centering\arraybackslash}m{15mm} | >{\centering\arraybackslash}m{12mm} |l|l|}
		\hline 
		Name & M($^{56}\rm Ni$) [$M_{\odot}$] & $t_0 [\rm days]$ &$\Sigma_{\rm Ni}t^2$ \newline [$10^{14}$gr cm$^{-2}$s$^2$]& M(IGE) [$M_{\odot}$]& M(IME) [$M_{\odot}$]& $M_{\rm tot}$ [$M_{\odot}$]& $E_K$ \newline $[10^{51} \rm ergs]$ & marker\\ 
		\hline 
		0.7 Ni: vary Ekin & 0.7 & 35 & 2.85, 3.06, 2.92, 2.78 & 0.1 & 0 & 1.21, 0.82, 1.04, 1.35 & 1.5, 0.5, 1.0, 2.0 & o blue\\ 
		\hline 
		vary $t_0$ & 0.7 & 25, 30, 40, 45 & 1.53, 2.11, 3.65, 4.42 & 0.1 & 0 & 0.95, 1.08, 1.34, 1.47 & 1.5 & o cyan\\ 
		\hline 
		const $\rho$ vary Ekin & 0.7 & 35 & 2.97, 2.47, 2.28, 2.16 & 0.1 & 0 & 0.92, 1.26, 1.52, 1.74 & 0.5, 1.0, 1.5, 2.0 & o green\\ 
		\hline 
		Ni-C vary mix$^a$ & 0.7 & 35 & 3.28, 3.23, 3.17, 3.08 & 0 & 0 & 1.13, 1.14, 1.14, 1.15 & 1.5 & o red\\ 
		\hline 
		outer composition & 0.7 & 35 & 2.85 & 0.1 & 0$^b$, 0.3 & 1.21 & 1.5 & o magenta\\ 
		\hline 
		0.3 Ni: vary Ekin & 0.3 & 35 & 2.03, 2.32, 2.14, 2.04 & 0.1 & 0 & 1.07, 0.70, 0.91, 1.21 & 1.5, 0.5, 1.0, 2.0 & $\square$ blue\\ 
		\hline 
		0.3 Ni: vary $t_0$ & 0.3 & 25, 30, 40, 45 & 1.14, 1.58, 2.68, 3.16 & 0.1 & 0 & 0.82, 0.95, 1.20, 1.32 & 1.5 & $\square$ cyan\\ 
		\hline 
		0.3 Ni: const $\rho$ vary Ekin  & 0.3 & 35 & 1.84, 1.61, 1.44, 1.36 & 0.1 & 0 & 0.87, 1.21, 1.47, 1.69 & 0.5, 1.0, 1.5, 2.0 & $\square$ green\\ 
		\hline 
		0.3 Ni: Ni-C vary mix$^a$ & 0.3 & 35 & 2.64, 2.45, 2.23, 1.95 & 0.1 & 0 & 0.99, 1.00, 1.02, 1.05 & 1.5 & $\square$ red\\ 
		\hline 
		0.3 Ni: outer composition & 0.3 & 35 & 2.03 & 0.1 & 0$^b$, 0.3 & 1.07 & 1.5 & $\square$ magenta\\ 
		\hline 
		0.1 Ni: vary Ekin & 0.1 & 35 & 1.87, 2.23, 2.01, 1.76 & 0 & 0 & 0.89, 0.54, 0.74, 1.02 & 1.5, 0.5, 1.0, 2.0 & x blue\\ 
		\hline 
		0.1 Ni: vary $t_0$ & 0.1 & 25, 30, 40, 45 & 1.07, 1.44, 2.30, 2.82 & 0 & 0 & 0.65, 0.77, 1.01, 1.13 & 1.5 & x cyan\\ 
		\hline 
		0.1 Ni: const $\rho$ vary Ekin  & 0.1 & 35 & 1.34, 1.16, 1.05, 0.98 & 0 & 0 & 0.83, 1.17, 1.44, 1.66 & 0.5, 1.0, 1.5, 2.0 & x green\\ 
		\hline 
		0.1 Ni: Ni-C vary mix$^a$ & 0.1 & 35 & 1.16, 1.16, 0.92, 0.70 & 0.1, 0, 0, 0 & 0 & 0.99, 0.94, 0.98, 1.02 & 1.5 & x red\\ 
		\hline 
		0.1 Ni: outer composition & 0.1 & 35 & 1.88 & 0 & 0$^b$, 0.3 & 0.89 & 1.5 & x magenta\\ 
		\hline 
		Kasen '07-like series & 0.6, 0.5, ..., 0.1 & 35 & 2.44, 2.03, 1.63, 1.22, 0.81, 0.41 & 0.1 & 0.1, 0.2, ..., 0.6$^c$ & 1.21 & 1.5 & v black\\ 
		\hline 
		Kasen '07-like no IMEs & 0.6, 0.5, ..., 0.1 & 35 & 2.44, 2.03, 1.63, 1.22, 0.81, 0.41 & 0.1 & 0$^d$ & 1.21 & 1.5 & v red\\ 
		\hline 
		Kasen '07-like no Fe & 0.6, 0.5, ..., 0.1 & 35 & 2.81, 2.34, 1.87, 1.40, 0.94, 0.47 & 0 & 0.1, 0.2, ..., 0.6$^c$ & 1.13 & 1.5 & v green\\ 
		\hline 		
		
	\end{tabular} 
	\flushleft
	\footnotesize{$^a$boxcar mixing iterations: 0, 20, 50, 100 \\
		$^b$C/O mix in outer layer \\
		$^c$Lowering the nickel homogeneously while replaceing it with IMEs \\
		$^d$Replacing the nickel homogeneously with carbon rather than IMEs}
\end{table*}

This work also addressed ejecta from different explosion scenarios: Central detonation of sub Chandrasekhar WDs from \cite{s+10} and delayed detonations of Chandrasekhar mass WDs from \cite{dbhk14}, both in spherical symmetry and for which radiative transfer simulations were presented. Along with these, two dimensional ejecta resulting form the direct collision of sub Chandrasekhar WDs from \cite{kk13} were discussed. The main physical parameters for these various ejecta are detailed in table~\ref{table:ejecta_table_literature}.

\begin{table*}
	\caption{ejecta from various models in literature discussed in the text.}
	\label{table:ejecta_table_literature}
	\begin{tabular}{|c|c|c|c|c|c|}
		\hline 
		Reference & Name & M($^{56}\rm Ni$) [$M_{\odot}$]& $t_0 [\rm days]$ &$\Sigma_{\rm Ni}t^2$ [$10^{14}$gr cm$^{-2}$s$^2$]& $M_{\rm tot}$ [$M_{\odot}$] \\ 
		\hline 
		Sim '10 & & 0.81 & 34 & 3.11 & 1.14 \\
		\hline
		Sim '10 & & 0.56 & 35 & 2.92 & 1.06 \\
		\hline
		Sim '10 & & 0.30 & 37 & 2.45 & 0.97 \\
		\hline
		Sim '10 & & 0.07 & 38 & 1.01 & 0.88 \\
		\hline
		Dessart '14 & DDC0 & 0.86 & 38 & 3.14 & 1.4 \\
		\hline
		Dessart '14 & DDC6 & 0.72 & 40 & 2.94 & 1.4 \\
		\hline
		Dessart '14 & DDC10 & 0.62 & 41 & 2.74 & 1.4 \\
		\hline
		Dessart '14 & DDC15 & 0.51 & 42 & 2.49 & 1.4 \\
		\hline
		Dessart '14 & DDC17 & 0.41 & 43 & 2.27 & 1.4 \\
		\hline
		Dessart '14 & DDC20 & 0.30 & 46 & 2.10 & 1.4 \\
		\hline
		Dessart '14 & DDC22 & 0.20 & 51 & 2.07 & 1.4 \\
		\hline
		Dessart '14 & DDC25 & 0.12 & 58 & 2.26 & 1.4 \\
		\hline				
		Kushnir '13 & 0.5-0.5 & 0.11 & 38 & 1.33 & 1.0 \\
		\hline				
		Kushnir '13 & 0.55-0.55 & 0.22 & 39 & 2.20 & 1.1 \\
		\hline				
		Kushnir '13 & 0.6-0.6 & 0.32 & 40 & 2.65 & 1.2 \\
		\hline				
		Kushnir '13 & 0.64-0.64 & 0.41 & 40 & 2.82 & 1.28 \\
		\hline				
		Kushnir '13 & 0.7-0.7 & 0.56 & 39 & 2.87 & 1.4 \\
		\hline				
		Kushnir '13 & 0.8-0.8 & 0.74 & 37 & 2.74 & 1.6 \\
		\hline				
		Kushnir '13 & 0.9-0.9 & 0.78 & 39 & 2.93 & 1.8 \\
		\hline				
		Kushnir '13 & 1.0-1.0 & 1.25 & 39 & 3.57 & 2.0 \\
		\hline				
		Kushnir '13 & 0.6-0.5 & 0.27 & 38 & 2.25 & 1.1 \\
		\hline				
		Kushnir '13 & 0.7-0.5 & 0.26 & 42 & 2.94 & 1.2 \\
		\hline				
		Kushnir '13 & 0.7-0.6 & 0.38 & 41 & 2.88 & 1.3 \\
		\hline				
		Kushnir '13 & 0.8-0.5 & 0.29 & 44 & 2.98 & 1.3 \\
		\hline				
		Kushnir '13 & 0.8-0.6 & 0.38 & 42 & 3.06 & 1.4 \\
		\hline				
		Kushnir '13 & 0.8-0.7 & 0.48 & 40 & 2.93 & 1.5 \\
		\hline				
		Kushnir '13 & 0.9-0.5 & 0.69 & 42 & 4.59 & 1.4 \\
		\hline				
		Kushnir '13 & 0.9-0.6 & 0.50 & 42 & 3.11 & 1.5 \\
		\hline				
		Kushnir '13 & 0.9-0.7 & 0.51 & 42 & 3.09 & 1.6 \\
		\hline				
		Kushnir '13 & 0.9-0.8 & 0.54 & 41 & 2.76 & 1.7 \\
		\hline				
		Kushnir '13 & 1.0-0.5 & 0.82 & 39 & 3.49 & 1.5 \\
		\hline				
		Kushnir '13 & 1.0-0.6 & 0.88 & 41 & 3.80 & 1.6 \\
		\hline				
		Kushnir '13 & 1.0-0.7 & 0.83 & 43 & 4.26 & 1.7 \\
		\hline				
		Kushnir '13 & 1.0-0.8 & 0.81 & 42 & 3.53 & 1.8 \\
		\hline				
		Kushnir '13 & 1.0-0.9 & 1.0 & 40 & 3.44 & 1.9 \\
		\hline
		
	\end{tabular} 
	
\end{table*}

\section{B-V Break Times in Observed SNIa Sample}\label{sec:appendixC}
In this work we focus on the WLR as is observed in the correlation between the light curve luminosity and the break time in the B-V curve. In addition, we show the bolometric WLR using the gamma-ray escape time. We therefore consider SNIa for which both B-V and bolometric light curves are available. The sample used is detailed in table~\ref{table:sample}. The sample was collected with the aim to cover the observed range (but not the distribution) of \nickel masses and is somewhat arbitrary.
The table contains values of $t_0$ and \nickel mass for each SNIa, as inferred from the bolometric lightcurve by the method detailed in paper I. The break time in the B-V curve, tbv, was determined in this work, as the intersection of two fitted straight lines, as explained in \S~\ref{sec:tbv_robust}. The observed B-V curves are found as function of maximal B time. In order to translate this to the time since explosion we use the recipe for rise time of the B band light curve given in equation 6 of \cite{s+14}: $t_{\rm R,B}=17.5-5(\Delta m_{15,B}-1.1)$ days, where the values of \dmft are taken from the literature, and given in the table as well.

\begin{table*}
	\caption{Observed SN sample. The values of $t_0$, \nickel, and tbv were obtained in this work using the bolometric and B-V light curves as well as the values of \dmft from the literature.}
	\label{table:sample}
	\begin{tabular}{|c|c|c|c|c|c|c|l|l|}
		\hline 
		name & $t_0$ [days] & M($^{56}\rm Ni$) [$M_{\odot}$]& tbv [days]& \dmft & ref for Bolometric & ref for B-V$^a$ \\ 
		\hline 
		1991T & 41 & 0.75 & 46.2 & 0.94 & \cite{strit05} & \cite{p+92, phil93} \\ 
		\hline 
		1991bg & 34 & 0.065 & 23.2 & 1.88 & \cite{strit05} & \cite{t+96, phil93} \\ 
		\hline 
		1992A & 33 & 0.22 & 36.4 & 1.47 & \cite{strit05} & \cite{g+04, h+96} \\ 
		\hline 
		1994D & 31 & 0.47 & 41.6 & 1.32 & \cite{strit05} & \cite{t+96, h+96} \\ 
		\hline 
		1994ae & 37 & 0.67 & 52.5 & 0.86 & \cite{strit05} & \cite{r+99} \\ 
		\hline 
		1995D & 42 & 0.48 & 48.5 & 0.99 & \cite{strit05} & \cite{r+99} \\ 
		\hline 
		1995ac & 40 & 0.76 & 48.5 & 1.01  & \cite{strit05} & \cite{r+99} \\ 
		\hline 
		1995al & 40 & 0.43 & 48.5 & 1.00 & \cite{strit05} & \cite{r+99} \\ 
		\hline 
		1998aq & 35 & 0.5 & 46.8 & 1.05 & \cite{strit05} & \cite{g+04, r+05} \\ 
		\hline 
		1998de & 34 & 0.06 & 24.1 & 1.95 & \cite{strit05} & \cite{mod+01} \\ 
		\hline 
		1999by & 35 & 0.07 & 22.9 & 1.9 & \cite{strit05} & \cite{g+04} \\ 
		\hline 
		2000cx & 30 & 0.3 & 46.7 & 0.93 & \cite{strit05} & \cite{can+03} \\ 
		\hline 
		2001el & 37 & 0.31 & 43.1 & 1.13 & \cite{strit05} & \cite{kris+03} \\ 
		\hline 
		2005cf & 40 & 0.54 & 45.5 & 1.07 & \cite{w+09} & \cite{w+09} \\ 
		\hline 
		2003du & 37 & 0.5 & 48.2 & 1.02 & \cite{s+07} &  \cite{s+07} \\ 
		\hline 
		2011fe & 40 & 0.43 & 43.3 & 1.21 & \cite{m+15} & \cite{rs12} \\ 
		\hline 
		2007on & 33 & 0.19 & 26.1 & 1.96 & \cite{phil12} & \cite{gall+18} \\ 
		\hline 
		2005ke & 35 & 0.075 & 25.6 & 1.77 & \cite{phil12} & \cite{burns14} \\ 
		\hline 
	\end{tabular} 
	\flushleft
	\footnotesize{$^a$where two references are cites, the B-V curve was taken from the first reference while \dmft from the second reference}
\end{table*}

\clearpage

\bibliographystyle{mnras}

\bsp
\label{lastpage}
\end{document}